\documentclass[prd,nofootinbib,preprint,superscriptaddress]{revtex4-1}
\usepackage{amsmath}
\usepackage{graphics}
\usepackage{epstopdf}
\usepackage{graphicx}
\usepackage{subfigure}

\makeatletter
\renewcommand{\fnum@table}{\textbf{\tablename~\thetable}}
\renewcommand{\fnum@figure}{\textbf{\figurename~\thefigure}}
\makeatother

\newcommand {\be}{\begin{equation}}
\newcommand {\ee}{\end{equation}}
\newcommand {\ba}{\begin{eqnarray}}
\newcommand {\ea}{\end{eqnarray}}

\begin{document}


\vspace*{10mm}

\title{Sensitivities of future solar neutrino observatories to NSI \vspace*{1.cm} }

\author{\bf Pouya Bakhti}
\affiliation{Institute for
	research in fundamental sciences (IPM), PO Box 19395-5531, Tehran,
	Iran}
\author{\bf Meshkat Rajaee}
\affiliation{Institute for
	research in fundamental sciences (IPM), PO Box 19395-5531, Tehran,
	Iran}

\begin{abstract}
  \vspace*{.5cm}
  
We study the matter effect caused by non-standard neutrino interactions (NSI) in
the future solar  neutrino experiments, DUNE, HK and  MICA. 
The upcoming reactor experiment, JUNO  is expected to provide the most precise measurements   of solar neutrino oscillation parameters and is going to open up the era of sub-percent precision in the leptonic mixing sector of the Standard Model (SM). Considering JUNO can measure $\Delta m ^2  _{21}$ and $\theta_{12}$ by  sub-percent precision and assuming SM as the null hypothesis,
 we study the possibility to constrain NSI parameters by the future solar neutrino
experiments such as  DUNE, HK and MICA.
For this purpose, we study the effect of NSI on solar neutrino propagation  in the
Sun and  Earth and  explore the dependence of the day-night asymmetry on the NSI parameters.    We also study the effect of NSI at the water Cerenkov detector on the simulated data for these experiments.

\end{abstract}

\maketitle

\section{Introduction}

Neutrino oscillation is well established by the data from a plethora of neutrino experiments using solar, atmospheric, reactor and accelerator neutrino experiments over the last two decades \cite{data}.
In the standard three flavor neutrino oscillation framework,
there are three mixing angles $\theta_{12}$, $\theta_{13}$, $\theta_{23}$, two mass-squared differences $\Delta m^2_{31}$, $\Delta m^2_{21}$ and one Dirac type CP phase $\delta_{CP}$.
Most of the oscillation parameters have been measured with fairly good precision \cite{copa,salas,esteban}; However, there are some unknown quantities, namely, the value of the Dirac CP phase $\delta_{\rm CP}$, mass ordering and the octant of $\theta_{23}$. To determine the unknown neutrino oscillation parameters, experiments
with high statistics, such as JUNO \cite{Djurcic:2015vqa},
T2HK \cite{t2hk}, and
DUNE \cite{dune},
have been proposed.

The framework of non-standard neutrino interaction (NSI) provides one model-independent way to extend the standard model to explain neutrino mass and to quantify new physics in the neutrino sector.
NSI was explored as a solution to
the solar neutrino problem \cite{ns}, and their impact on the oscillations of solar neutrinos \cite{so}, atmospheric neutrinos
\cite{at}, and accelerator neutrinos \cite{ac} have been explored in the literature. Moreover, several consequences of NSI to DUNE were also explored in \cite{masud}. Moreover, there is a tension between the mass-squared difference obtained from the
solar neutrino observations and the one from the KamLAND experiment.
As studied in \cite{tension}, one proposed solution is the sterile
neutrino oscillation with the mass-squared difference
of order of O(10$^{-5}$) eV$^2$, which is so-called Super light Sterile Neutrino Scenario (SSNS).
Another possibility is that the tension can be resolved by introducing the flavor-dependent
NSI in neutrino propagation \cite{Gonzalez-Garcia:2013usa,wolf}.

The Jiangmen Underground Neutrino Observatory (JUNO) experiment is a future reactor experiment with a baseline of 50 km. The main purpose of JUNO is to determine the mass ordering and it will measure $\Delta m^2_{21}$ and $\theta_{12}$ to the percent level \cite{jun}. However, JUNO is not sensitive to the NSI parameters due to its low neutrino energy \cite{Bakhti:2014pva}. Ref. \cite{Bakhti:2013ora} have also studied the potential of JUNO to test SSNS. Since solar neutrino oscillation probabilities are strongly dependent on the NSI parameters due to the matter effect, with precise measurement of $\Delta m^2_{21}$ and $\theta_{12}$ with JUNO in the presence of NSI, it is crucial to investigate how well the future solar neutrino observatories can constrain non-standard neutrino interaction.
In this work, we consider future solar neutrino experiment, DUNE and HK in addition to the proposed solar neutrino experiment, MICA. We explore the potential of these experiments in resolving standard parameter degeneracies in the presence of NSI.
It is possible to study the effect of NSI of solar neutrinos with the matter in the
Sun and Earth \cite{Gonzalez-Garcia:2013usa,Liao:2017awz}.
Considering the day-night asymmetry of solar neutrino, we also study the dependence of the day-night asymmetry on the NSI parameters.
For simplicity, we assume the same NSI couplings to electron, up quark and down quark.
As it is discussed in detail, assuming non-standard couplings to electrons will affect the electron-neutrino scattering cross-section and can lead to NSI at the HK and MICA detectors. In this paper, we explore the effect of NSI on the neutrino detection for HK and MICA detectors experiments which are water Cerenkov detectors \cite{Bolanos:2008km}.

The paper is organized as follows. In Sec.~II, we discuss the NSI Lagrangian and its effect on solar neutrino oscillation. In Sec.~III, we discuss the details of the different experiment and our simulation. In Sec.~IV, we present our results. We summarize our results in Sec.~V.

\section{Non-Standard Neutrino Interaction\label{theo}}

Neutral current (NC) NSI can be written as an effective four fermion operator

\begin{equation}
\label{eq:def}
\mathcal{L}_\text{NSI} =
- 2\sqrt{2} G_F \epsilon_{\alpha\beta}^{fP}
(\bar\nu_{\alpha} \gamma^\mu \nu_{\beta})
(\bar{f} \gamma_\mu P f) \,,
\end{equation}
where $f$ is a charged fermion, $P=(L,R)$ and
$\epsilon_{\alpha\beta}^{fP}$ are dimensionless parameters encoding the
deviation from standard interactions and $G_F$ is the Fermi coupling constant.
Constraints on $\epsilon_{\alpha\beta}$
have been discussed in many references. For instance, there are bounds from atmospheric neutrinos \cite{garcia,lipari,valle,bayo},
from $e^+ e^-$ colliders \cite{Berezhiani},
from the compilation of various neutrino data \cite{Davidson,Biggio},
from solar neutrinos \cite{Friedland,Miranda,Palazzo},
from $\nu_e e$ or $\bar{\nu}_e e$ scatterings \cite{Miranda},
from solar, reactor and accelerator neutrinos \cite{moura,Escrihuela}. In addition, NSI has been studied in the context of long-baseline
experiments \cite{Adhikari,masud,deGouvea}

NSI can be induced by the new
physics beyond the standard model, by integrating out the heavier mediator fields which can generate the dimension-6 \cite{six} and dimension-8 \cite{eight}
effective operators. For a detailed review see Refs. \cite{ohlsson,miranda}

The neutral current NSI affect the neutrino oscillation in matter via forward elastic scattering. NC NSI also can affect the neutrino detection via neutrino electron scattering. In this work, we consider the effect of NSI on solar neutrinos for three cases: (i) through their propagation in the sun, (ii) through their propagation in the earth and (iii) and by water Cerenkov detectors.

\subsection{Propagation of Neutrinos in the Sun in the Presence of NSI\label{theo}}

In the flavor basis, the flavor change of neutrinos through the propagation can be written as
\begin{equation}
i\, \frac{d}{dx}\psi_\nu = H \psi_\nu
\end{equation}
where the total Hamiltonian includes the vacuum effect, standard matter effect or MSW effect and NSI matter effect
\begin{equation}
H = H_\mathrm{vac} + H_\mathrm{mat}^\mathrm{MSW} + H_\mathrm{mat}^\mathrm{NSI}
\label{H_tot}
\end{equation}
The vacuum term includes six parameters, $\Delta m^2_{21}$, $\Delta m^2_{31}$, $\theta_{12}$, $\theta_{13}$, $\theta_{23}$, $\delta_{CP}$ and is given by
\begin{equation}
H_{vac} = U\text{diag}
\left(0, \frac{\Delta m^2_{21}}{2E_\nu} , \frac{\Delta m^2_{31}}{2E_\nu}\right)
U^\dagger
\end{equation}
where $U$ is the standard Pontecorvo-Maki-Nakagawa-Sakata mixing matrix, $U = R_{23}\Gamma_\delta R_{13}\Gamma_\delta^\dagger R_{12}$, where $R_{ij}$ represents a real rotation by an angle $\theta_{ij}$ in the $ij$ plane, $\Gamma_\delta=\text{diag}(1,1,e^{i\delta})$. The Hamiltonian of standard matter effect is given by $H^\text{MSW}_\text{mat} =\sqrt{2}
G_F N_e \text{diag} (1, 0, 0)$, where $N_e$ is the number density of electron in the medium. Moreover, The NSI matter effect is given by
\begin{equation}
\label{eq:hmatNSI}
H_\mathrm{mat}^\mathrm{NSI}=
\sqrt{2} G_F \sum_{f=e,u,d} N_f
\begin{pmatrix}
\epsilon_{ee}^f & \epsilon_{e\mu}^f & \epsilon_{e\tau}^f
\\
\epsilon_{e\mu}^{f*} & \epsilon_{\mu\mu}^f & \epsilon_{\mu\tau}^f
\\
\epsilon_{e\tau}^{f*} & \epsilon_{\mu\tau}^{f*} & \epsilon_{\tau\tau}^f
\end{pmatrix} .
\end{equation}
It is possible to define NSI parameter in the medium
\begin{equation}
\label{eq:compoconst}
\epsilon_{\alpha\beta} \equiv
\sum_{f=e,u,d} \left< \frac{N_f}{N_e} \right> \epsilon_{\alpha\beta}^f
= \epsilon_{\alpha\beta}^e + Y_u\, \epsilon_{\alpha\beta}^u + Y_d\, \epsilon_{\alpha\beta}^d
\end{equation}
where $Y_\alpha$ is the ratio of averaged fermion number density over electron number density in the medium. In the sun $Y_u\approx2$ and $Y_d\approx1$ and in the earth $Y_u\approx Y_d \approx 3$.

Since $\frac{\Delta m^2_{31}}{E_\nu}\gg G_F N_e$ for solar neutrinos, it is possible to work on one mass dominate approximation, using $2\times2$ effective Hamiltonian as following

\begin{align}
\label{eq:hvacsol}
H_\text{vac}^\text{eff}
&= \frac{\Delta m^2_{21}}{4 E_\nu}
\begin{pmatrix}
-\cos 2\theta_{12} & \sin 2\theta_{12} \\
\hphantom{+} \sin 2\theta_{12} & \cos 2\theta_{12}
\end{pmatrix} ,
\\
\label{eq:hmatsol}
H_\text{mat}^\text{eff}
&= \sqrt{2} G_F N_e(r)
\begin{pmatrix}
c_{13}^2 & 0 \\
0 & 0
\end{pmatrix}
+ \sqrt{2} G_F \sum_f N_f(r)
\begin{pmatrix}
-\epsilon_D^{f\hphantom{*}} & \epsilon_N^f \\
\hphantom{+} \epsilon_N^{f*} & \epsilon_D^f
\end{pmatrix} .
\end{align}
The coefficients $\epsilon_D^f$ and $\epsilon_N^f$ are given with respect to the original
parameters $\epsilon_{\alpha\beta}^f$ as the following \cite{Gonzalez-Garcia:2013usa}
\begin{align}
\label{eq:epsD}
\epsilon_D^f =
-\frac{c_{13}^2}{2} \big( \epsilon_{ee}^f - \epsilon_{\mu\mu}^f \big)
+ \frac{s_{23}^2 - s_{13}^2 c_{23}^2}{2}
\big( \epsilon_{\tau\tau}^f - \epsilon_{\mu\mu}^f \big)
+ \text{Re}\left[ c_{13} s_{13}e^{i\delta} \big( s_{23} \, \epsilon_{e\mu}^f
+ c_{23} \, \epsilon_{e\tau}^f \big)
- \big( 1 + s_{13}^2 \big) c_{23} s_{23} \epsilon_{\mu\tau}^f \right]
\,,
\end{align}
\begin{align}
\label{eq:epsN}
\epsilon_N^f =
c_{13} \big( c_{23} \, \epsilon_{e\mu}^f - s_{23} \, \epsilon_{e\tau}^f \big)
+ s_{13} e^{-i\delta} \left[
s_{23}^2 \, \epsilon_{\mu\tau}^f - c_{23}^2 \, \epsilon_{\mu\tau}^{f*}
+ c_{23} s_{23} \big( \epsilon_{\tau\tau}^f - \epsilon_{\mu\mu}^f \big)
\right]\,.
\end{align}

Then the effective Hamiltonian can be diagonalized as~\cite{Liao:2015rma}
\begin{align}
U'=\left(\begin{array}{cc}
\cos\tilde{\theta}_{12} & \sin\tilde{\theta}_{12} e^{-i\phi}
\\
-\sin\tilde{\theta}_{12} e^{i\phi} & \cos\tilde{\theta}_{12}
\end{array}\right)\,,
\end{align}
where
\begin{align}
\tan2\tilde{\theta}_{12}=\frac{|\sin2\theta_{12}+2\hat{A}_E\epsilon_N|}{\cos2\theta_{12}-\hat{A}_E(c_{13}^2-2\epsilon_D)}\,,
\end{align}
and
\begin{align}
\phi=-\text{Arg}\left(\sin2\theta_{12}+2\hat{A}_E\epsilon_N\right)\,.
\end{align}
Thus, solar neutrino oscillation probability during the day is given by
\begin{equation}
P_D (E) = \frac{1}{2} {c_{13}^4}\left[1 + \cos 2 \theta_{12} \cos 2 \tilde{\theta}_{12}(E) \right] + s_{13}^4
\end{equation}

\subsection{Day-Night Asymmetry and NSI\label{theo}}
Due to loss of coherence, the solar neutrinos arrive
at the surface of the Earth as independent fluxes of the
mass eigenstate. Inside the Earth, the mass states oscillate in multi-layer medium and the oscillations proceed in low density regime which is quantified by small parameter
\begin{equation}
\epsilon \equiv \frac{2 V E}{\Delta m_{21}^2 }
\end{equation}
where $V(x) = \sqrt{2}G_F n_e(x)$ is the matter potential.

During the night, neutrinos travel a larger distance to reach the detector because of propagation through the Earth. The differences of survival probability during night and day is given by \cite{Bakhti,Liao:2015rma}
\begin{equation}
\Delta {P} (E, \eta) = P_N - P_D= \kappa(E) \left[\int_{0}^{L} \! dx \ V(x) \sin \phi^m ( L - x, E) + I_2 \right]
\end{equation}
where
\begin{equation}
\kappa (E) \equiv -\frac{1}{2}c_{13}^4 \cos 2 \tilde{\theta}_{12}^s \sin 2 \theta_{12} (\sin 2 \theta_{12} (c_{13}^2-2\epsilon_D^E)+2\cos 2 \theta_{12}\epsilon_N^E)
\end{equation}
and
\begin{equation}
I_2 = \frac{1}{2}\cos 2\theta_{12}\left (\int_{0}^{L} \! dx \ V(x) \cos \phi^m (L - x) \right)^2
\label{eq:secondod}
\end{equation}
and
\begin{equation}
\phi^m (L-x, E) \equiv \int_{x}^{L} \! \! d x \ \Delta_{21}^m(x)
\end{equation}
where
\begin{equation}
\Delta_{21}^m = \Delta_{21} \sqrt{(\cos 2\theta_{21}-(c_{13}^2-2\epsilon_D^E)a_{CC}^E)^2+|\sin2\theta_{12}+2a_{CC}^E\epsilon_N^E|^2 }\approx \Delta_{21} (1- c_{13}^2 \cos 2 \theta_{12} a_{CC}^E)
\end{equation}
where $a_{CC}^E\equiv 2 V(x) E/\Delta m_{21}^2$ and $\Delta_{21}\equiv\Delta m_{21}^2/4E$. As discussed in \cite{Bakhti} we can neglect $I_2$ in our calculations. For a constant density
\begin{equation}
\Delta P(E,\eta) = -\frac{1}{2}c_{13}^6 \cos 2 \bar{\theta}^\odot_{12}(E) \sin^2 2 \theta_{12}
\times
(\frac{a_{CC}}{1-c_{13}\cos(2\theta_{12})a_{CC}}(1-\cos (L\Delta_{21}(1-c_{13}\cos(2\theta_{12})a_{CC}))
\end{equation}
where $\eta$ is the nadir angle and
\begin{equation}
L=\cos\eta.
\end{equation}

Considering the effective resolution function $g(E_r, E)' = g(E_r, E) \sigma (E) f_B(E)$, where $\sigma (E)$ is the neutrino interaction cross section and $f_B(E)$ is the flux, and plugging expression for $\Delta P(E)$,
we have the following integral
\begin{eqnarray}
I_\Delta (E_r) \equiv \int dE~ g_\nu'(E_r, E) \Delta P(E) =
\int_0^L dx V(x) \int_0^{E_B} dE~ g_\nu'(E_r, E) \sin \phi^m ( L - x, E)
\label{eq:permut}
\end{eqnarray}
where we permuted integration over $x$ and $E$.
Let us introduce the attenuation factor $F(L-x)$ substituting the integral over
$E$ in Eq.~(\ref{eq:permut}) by
\begin{equation}
F(L-x) \sin \phi^m (L - x, E_r) = \int dE g_\nu'(E_r, E) \sin \phi^m (L - x, E).
\label{eq:attenf}
\end{equation}
In general, this equality cannot be satisfied,
but it is valid for special cases and under integral over $x$.
Then the integral (\ref{eq:permut}) becomes
\begin{equation}
I_\Delta (E_r) = \int dx V(x) F(L-x) \sin \phi^m (L - x, E_r).
\label{eq:permut1}
\end{equation}
For the ideal resolution, $g(E_r, E) = \delta (E_r - E)$, the Eq.~(\ref{eq:attenf}) gives
$F(L-x) = 1$ which means that attenuation is absent.

For the Gaussian energy resolution function the attenuation factor is given by
\begin{equation}
F(d)\simeq e^{-2\left({ d \over \lambda_{att} }\right)^2}
\label{eq:attfactorx}
\end{equation}
where
\begin{equation}
\lambda_{att} \equiv l_\nu \frac{E}{\pi \sigma_E}
\label{eq:attlength}
\end{equation}
is the attenuation length, and $l_\nu$ is the oscillation length in vacuum
\begin{equation}
l_\nu = \frac{4\pi E}{\Delta m_{21}^2}.
\label{eq:osclength}
\end{equation}
As can be seen in Eq.~\ref{eq:attfactorx}, for $d$ much larger than $\lambda_{att}$ (remote deep interiors), $F(d)$ goes to zero, while for $d$ and $\lambda_{att}$ at same order (the shallower interior), $F(d)$ becomes large, thus day-night asymmetry depends on the shallower interior more than deeper interior of the Earth.

Day-night asymmetry is defined as
\begin{equation} \label{day}
A_{ND}(E_r, \eta) \equiv \frac{N_N}{N_D}-1
\end{equation}
where

\begin{equation}
N_{D (N)} = A \int dE g_{\nu} (E^r , E) \sigma (E) f_B (E)  P (E) _{D (N)}
\end{equation}
where A is the factor which includes characteristics of detection.

The averaged over the year asymmetry is given by integrating ${A}_{DN}$ multiplied by the exposure (weight)
function $W(\eta)$ over nadir angle.
\begin{equation}
\bar{A}_{DN} = \int d \eta W (\eta) {A}_{DN} (\eta).
\end{equation}

\subsection{Neutral Current NSI Effect on Electron Neutrino Scattering\label{theo}}
The differential cross-section of neutrino electron scattering as a function of electron kinetic energy is well known and is given by
\begin{equation}
\label{eq:cs}
\frac{d\sigma}{dT} (E_\nu, T_e)=
{2 G_F^2 m_e \over \pi}
\left[(g_1)^2 + (g_2)^2\left(1-\frac{T_e}{E_{\nu}}\right)^2-
g_1 g_2 \frac{m_e T_e}{ E^2_{\nu}}\right],
\end{equation}
where within the standard model
\begin{equation}
g_1^{\nu_e}=g_2^{\bar{\nu}_e}=\frac{1}{2}+\sin^2\theta_W=0.73
\end{equation}
\begin{equation}
g_2^{\nu_e}=g_1^{\bar{\nu}_e}=g_2^{\nu_\mu}=g_1^{\bar{\nu}_\mu}=\sin^2\theta_W=0.23
\end{equation}
\begin{equation}
g_1^{\nu_\mu}=g_2^{\bar{\nu}_\mu}=-\frac{1}{2}+\sin^2\theta_W=-0.27.
\end{equation}

The total cross-section of neutrino electron scattering as a function of energy threshold and neutrino energy is given by

\begin{align}
\label{eq:cs}
\sigma(E_\nu,T_e^{th}) &= \frac{2G_F^2m_e}{\pi}\left[(g_1^2+g_2^2)(T_e^{max}-T_e^{th})- \left(g_2^2+g_1 g_2 \frac{m_e}{2E_\nu} \right) \right. \\
& \left. \left(\frac{T_e^{max^2}-T_e^{th^2}}{E_\nu} \right)+\frac{1}{3} g_2^2 \left(\frac{T_e^{max^3}-T_e^{th^3}}{E_\nu^2} \right) \right]
\end{align}

where
\begin{equation}
T_e^{max}(E_\nu)=\frac{2 E_\nu^2}{m_e+2E_\nu}
\end{equation}
Considering the neutral current NSI with electron $g_1^e$ and $g_2^e$ modifies as following \cite{Bolanos:2008km},
\begin{equation}
g_1^{e~NSI}=g_1^e+\epsilon^{eL}_{ee}
\end{equation}
\begin{equation}
g_2^{e~NSI}=g_2^e+\epsilon^{eR}_{ee}
\end{equation}

\section{ Details of Experiment and our Analysis}

We have considered solar neutrino detection with Hyper-Kamiokande, DUNE, and MICA experiments. For the statistical inferences we have considered ten years of data taking for each detector. Hyper-Kamiokande (HK) will detect the solar neutrinos by
neutrino-electron elastic scattering with 6.5~\rm{MeV} threshold \cite{Hyper-Kamiokande:2016dsw}. Considering 0.5 Mton fiduciary volume, we have calculated 200 events per day \cite{Liao:2015rma}. We have assumed the energy resolution of HK is $\sigma_E/E=0.15$. DUNE will have fiducial volume $40~$ kt liquid argon. On the other hand, DUNE can detect solar neutrinos via charged current interaction
\begin{equation}
\nu_e + ^{40 } Ar \rightarrow ^{40} K + e^-
\end{equation}
We consider the generic form of cross-sections for interaction with
nuclei
\begin{equation}
\sigma_{CC} (E) = A {p_e E_e},
\label{eq:crossect}
\end{equation}
where $A$ is a normalization factor (irrelevant for the relative excess),
$E_e = E_\nu - \Delta M$, $p_e$ is the electron momentum
and $\Delta M = 5.8~\rm{MeV}$ is the reaction threshold \cite{Ioannisian:2017dkx}.
Notice that only 9.7\% of $^8$B neutrinos have energy
$E_\nu$ $ >$ 11~\rm{MeV} but due to strong energy dependence (\ref{eq:crossect})
the corresponding fraction of the detected events is 0.9. We find that about 27000 events of $\nu_e$ will be detected
annually for $E_\nu > 11~$ MeV in a 40 kt liquid argon detector considering neutrino interaction with argon nuclei \cite{Ioannisian:2017dkx,Bakhti}. We have assumed the energy resolution of DUNE is $\sigma_E/E=0.07$.

MICA is a proposed detector that will be located at Amundsen-Scott South Pole station \cite{Boser:2013oaa},
in the same place as ICECUBE.
In our calculations we have taken the characteristics of MICA from Ref.~\cite{Boser:2013oaa},
10 Mton fiducial mass and 10~\rm{MeV} energy threshold for the kinetic energy
of the recoil electron. With these parameters,
we find that about 5$\times 10^5$ solar
$\nu e -$ scattering events are expected per year. We have assumed the energy resolution of MICA will be $\sigma_E/E=0.15$.

We have considered only solar boron neutrino flux from Ref.~\cite{Bahcall:1996qv}. For the analysis of solar neutrino detection, we have considered ten years of data taking. We have assumed solar neutrino parameters true value are $\Delta m^2_{21}=7.5\times10^{-5}$ and $\theta_{12}=33.5^\circ$. Our analysis shows that in the presence of NSI, $\Delta m^2_{21}$ and $\theta_{12}$ will be determined with a precision of 1.2$\times 10^{-6}~{\rm eV^2 }$ and 0.07$^\circ$ respectively with JUNO after 10 years of data taking. We have considered all the details of JUNO the same as given in Ref.~\cite{Bakhti:2013ora,Bakhti:2014pva}. Since these two parameters will be determined with very high precision, we fix these parameters in our analysis. We have assumed the PREM model \cite{prem} for the Earth structure to calculate the day-night asymmetry. In addition, we have assumed the same details of analysis as given in Ref.~\cite{Bakhti}. For the statistical inferences we have considered Asimov data set approximation, and the true model is the standard model or $\epsilon_{\alpha\beta}^f=0$. For statistical inferences for oscillation of the neutrinos in the sun, we have used chi squared method. We have neglected the matter effect in the earth or day-night asymmetry for constraining the parameters from the effect of NSI on the oscillation in the sun. For statistical errors of day-night asymmetry is calculated with the assumption of Gaussian distribution of errors. Moreover, in all of our calculation we have assumed $\epsilon_{\alpha\beta}^e=\epsilon_{\alpha\beta}^u=\epsilon_{\alpha\beta}^d=\epsilon_{\alpha\beta}$.

\section{Results}

\begin{figure}[h]
\hspace{0cm}
\includegraphics[width=0.45\textwidth, height=0.35\textwidth]{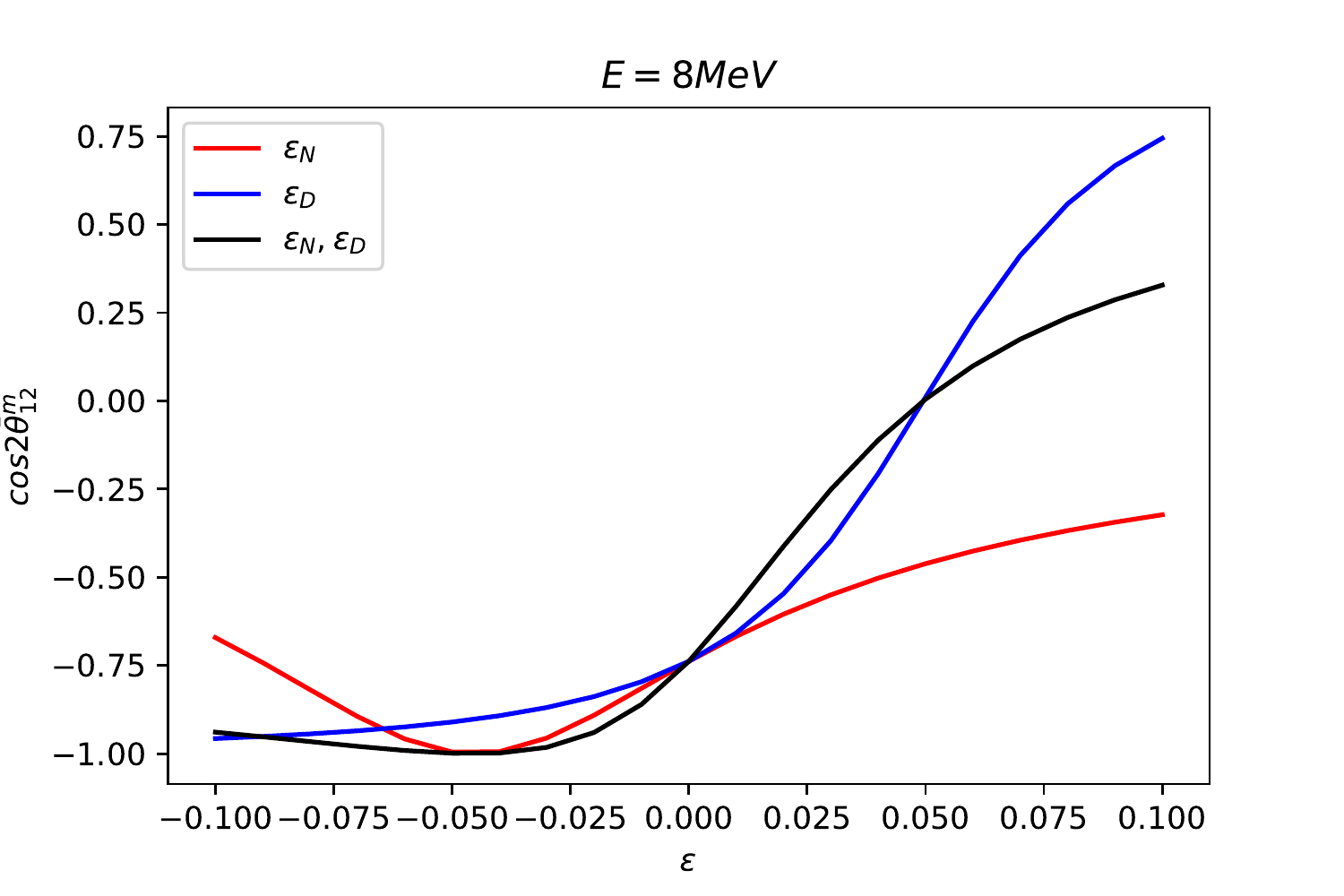}
\hspace{0cm}
\includegraphics[width=0.45\textwidth, height=0.35\textwidth]{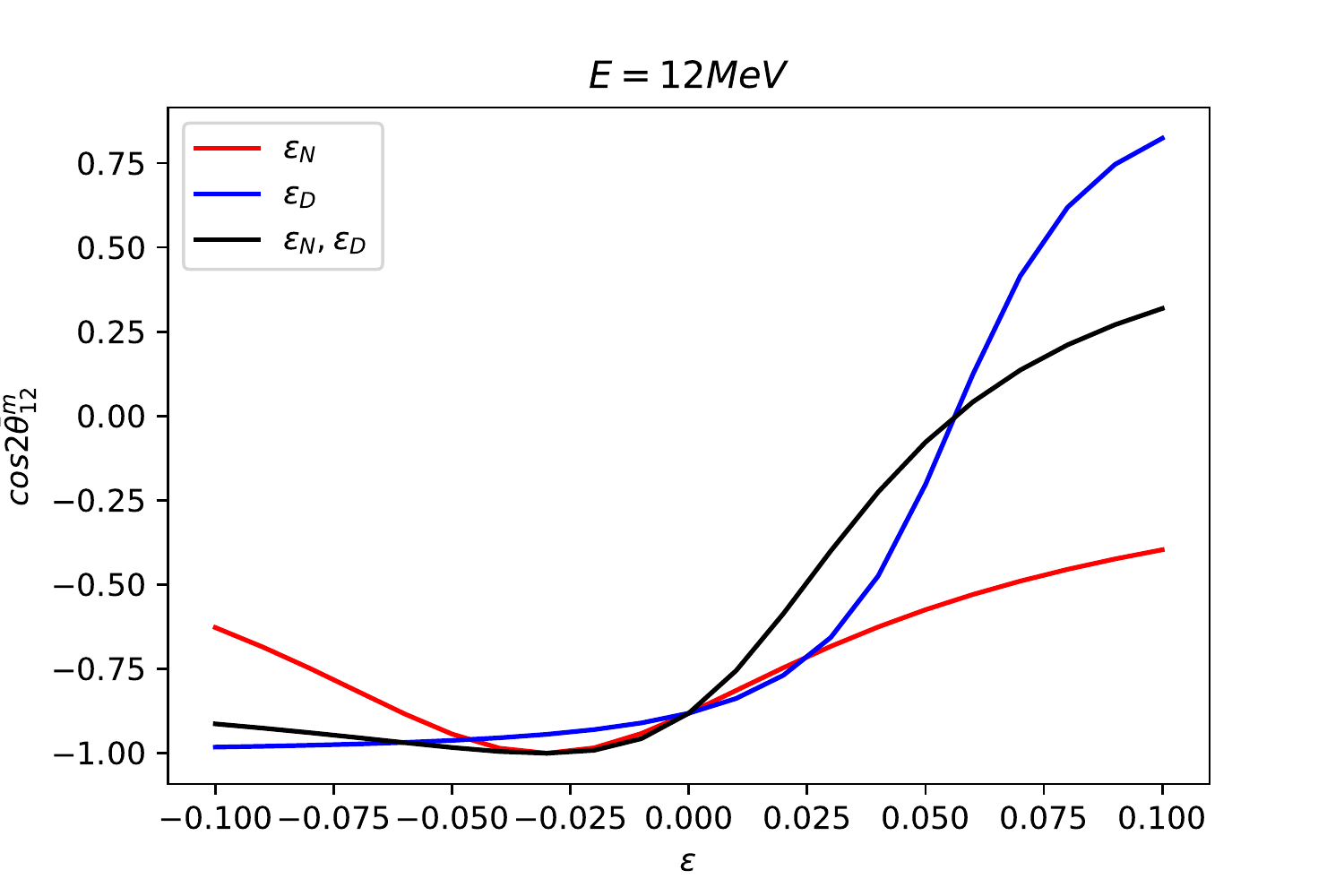}
\hspace{0cm}
\includegraphics[width=0.45\textwidth, height=0.35\textwidth]{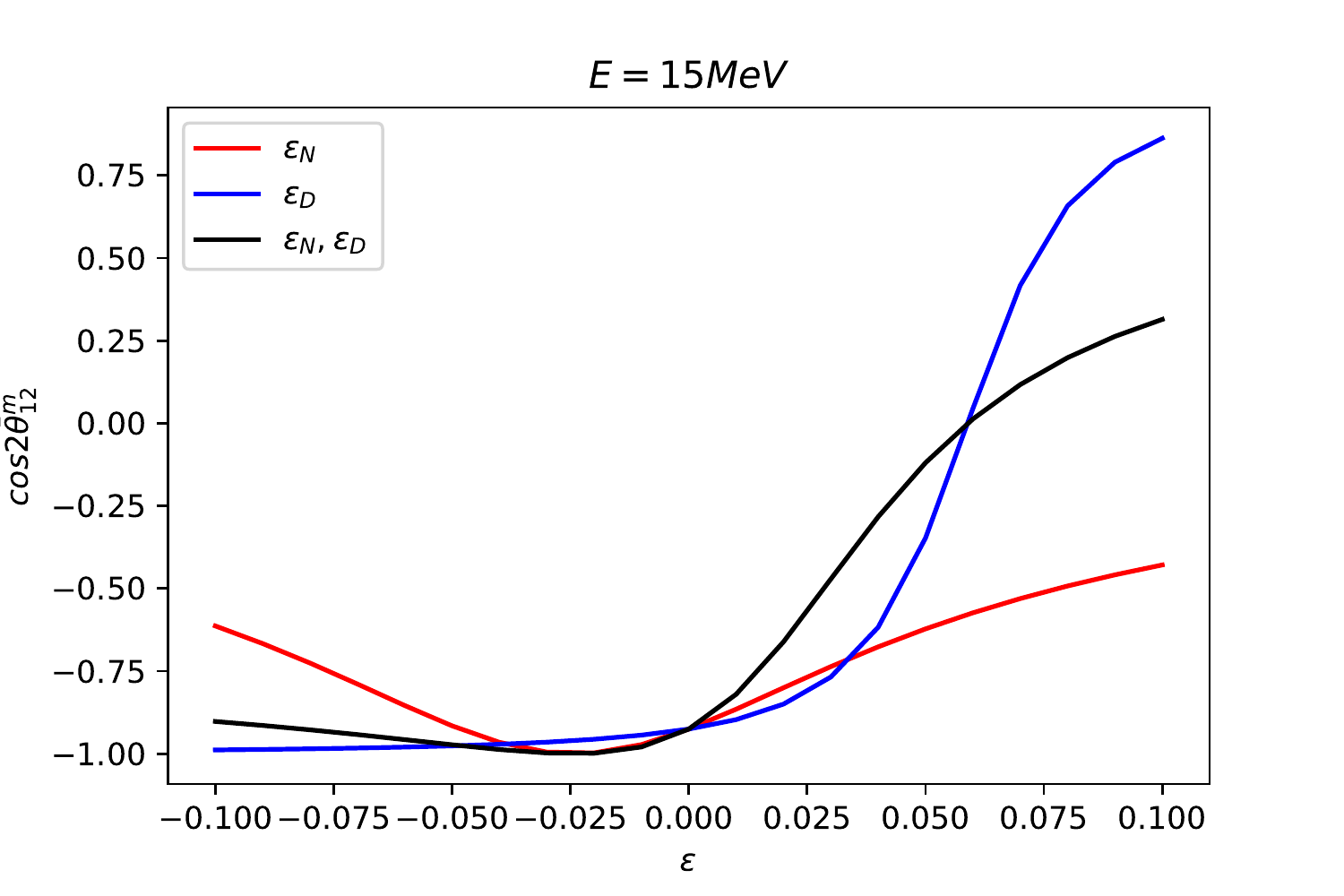}
\caption[...]{
$\cos 2 \tilde{\theta}_{12}$ as a function of  $\epsilon$ is plotted for three different neutrino energies of 8, 12, 15 MeV.
We have assumed $\epsilon_e = \epsilon_u = \epsilon_d=\epsilon_{\alpha\beta}$. The blue, red and black curves are plotted assuming $\epsilon_D \neq 0 $, $\epsilon_N \neq 0 $ and $\epsilon_D =\epsilon_N \neq 0 $, respectively.
\label{cos2t_sun1}
}
\end{figure}

\begin{figure}[h]
\hspace{0cm}
\includegraphics[width=0.55\textwidth, height=0.4\textwidth]{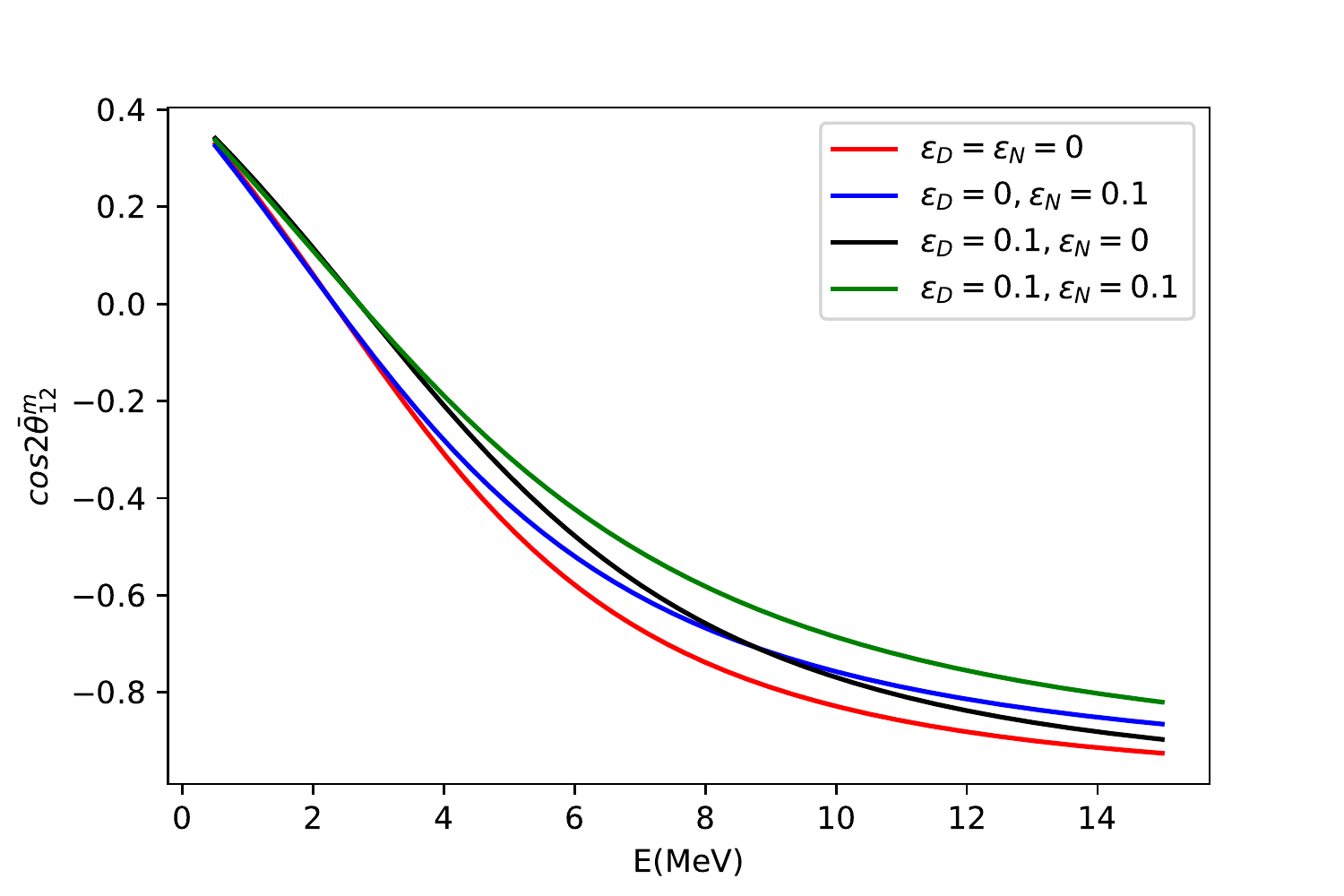}
\caption[...]{
$\cos 2 \tilde{\theta}_{12}$ versus energy is plotted assuming $\epsilon=0.1$.
\label{cos2t_sun2}
}
\end{figure}

\begin{figure}[h]
\hspace{0cm}
\includegraphics[width=0.45\textwidth, height=0.35\textwidth]{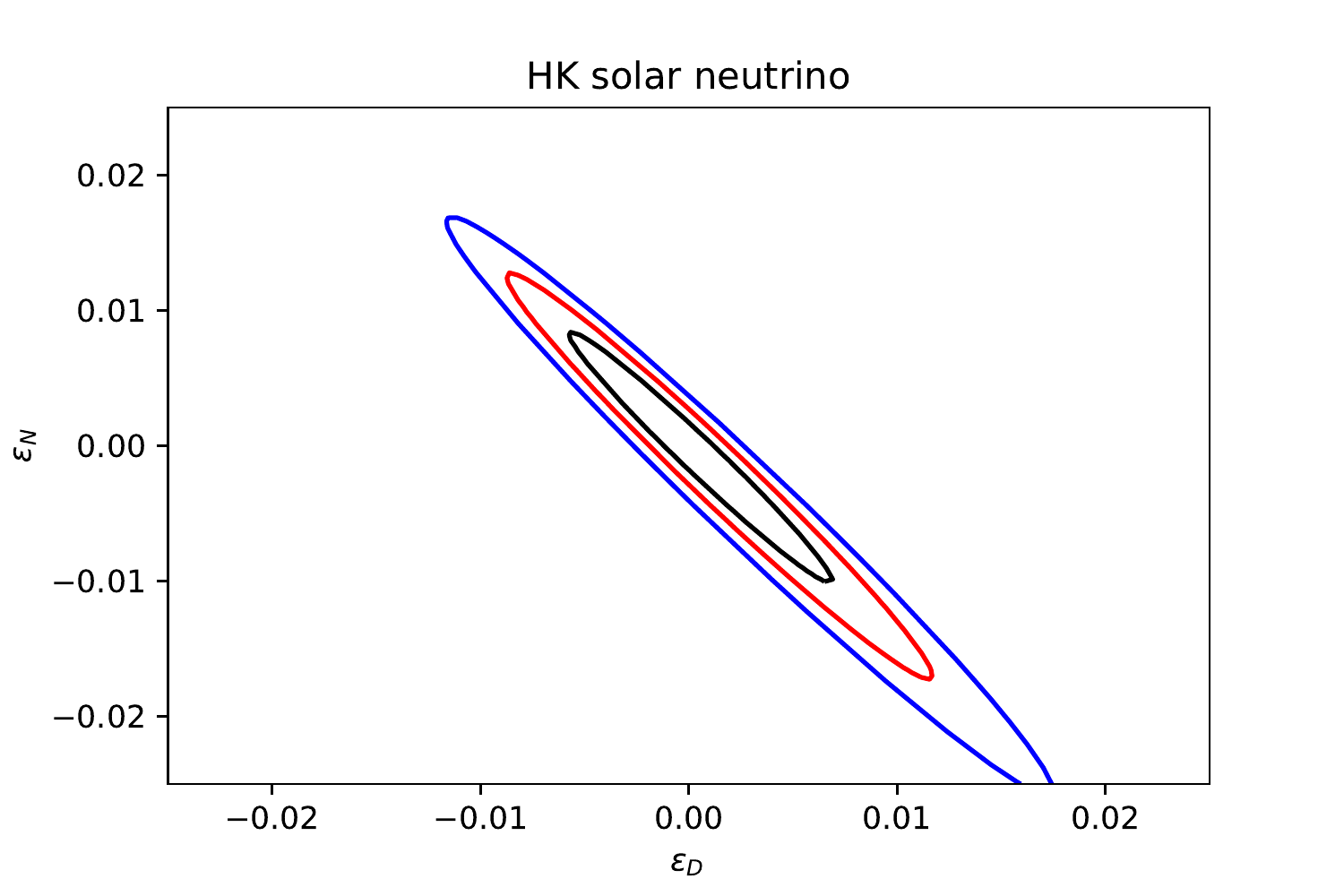}
\hspace{0cm}
\includegraphics[width=0.45\textwidth, height=0.35\textwidth]{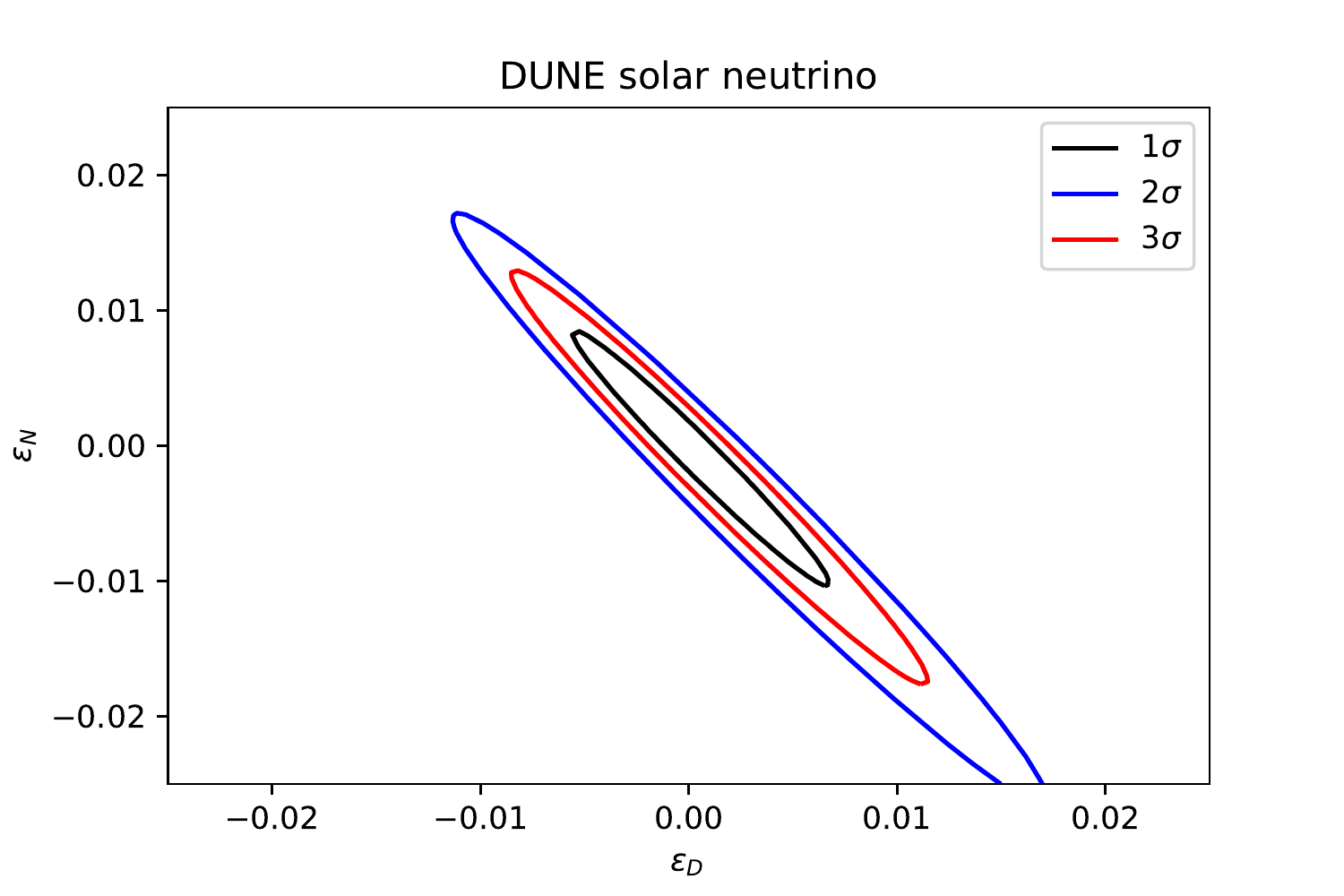}
\hspace{0cm}
\includegraphics[width=0.45\textwidth, height=0.35\textwidth]{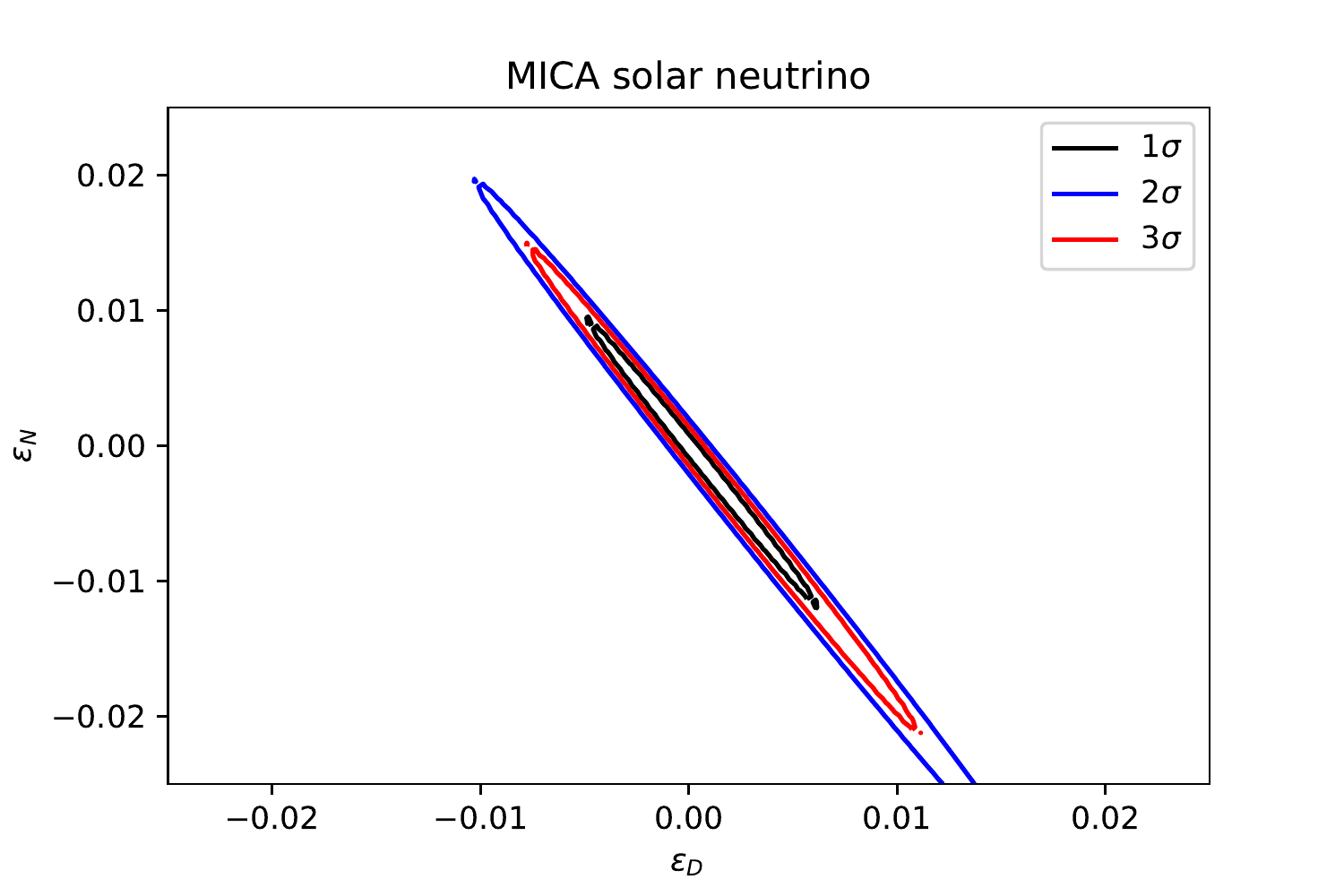}
\hspace{0cm}
\includegraphics[width=0.45\textwidth, height=0.35\textwidth]{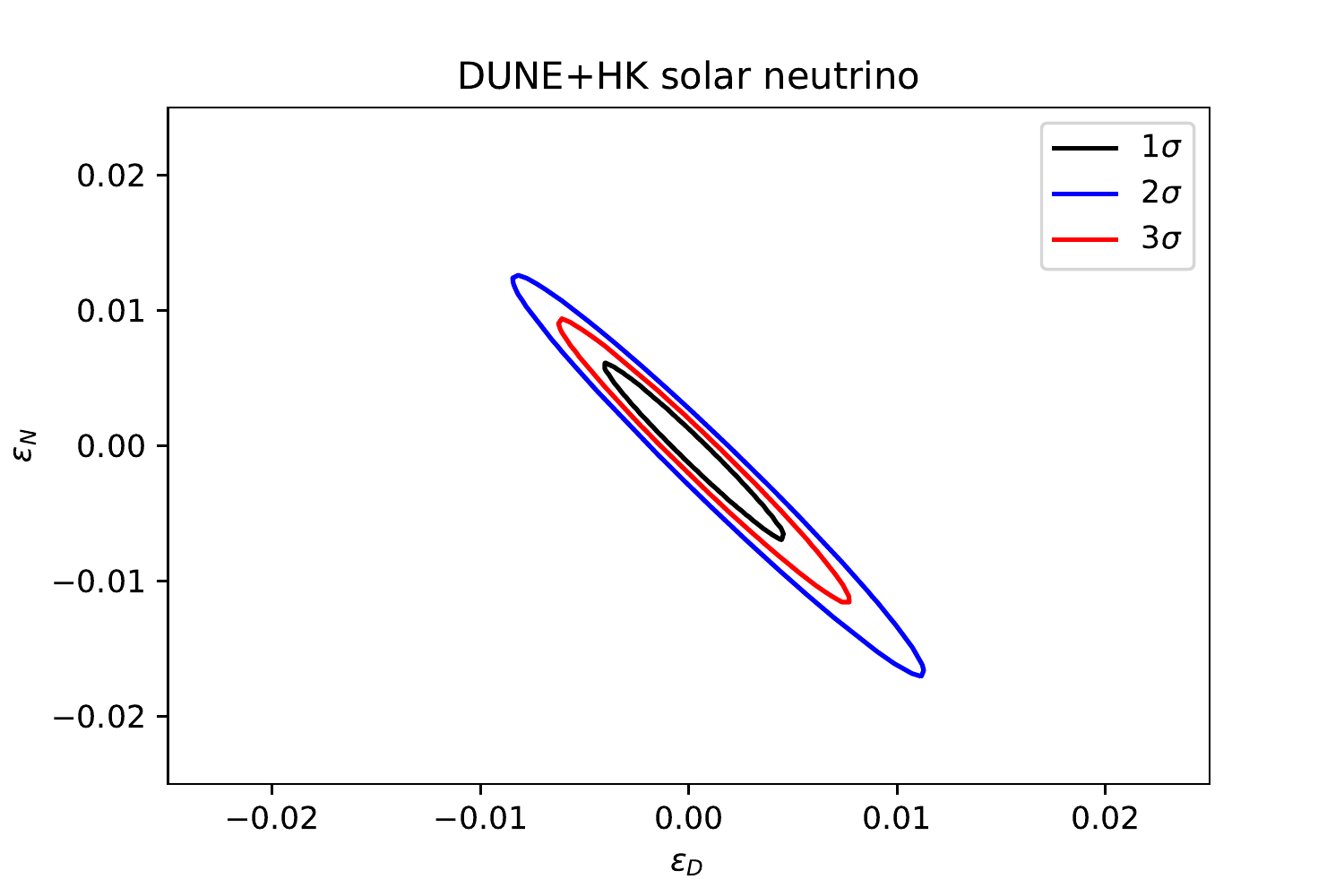}
\caption[...]{
1$\sigma$, 2$\sigma$ and 3$\sigma$ allowed regions for $\epsilon_D$
versus $\epsilon_N$ at DUNE, HK, MICA, and combination of DUNE and HK assuming $\Delta m^2_{21}=7.5\times10^{-5}$ and $\theta_{12}=33.5^\circ$.
\label{constraints_eps}
}
\end{figure}

\begin{figure}[h]
\hspace{0cm}
\includegraphics[width=0.45\textwidth, height=0.35\textwidth]{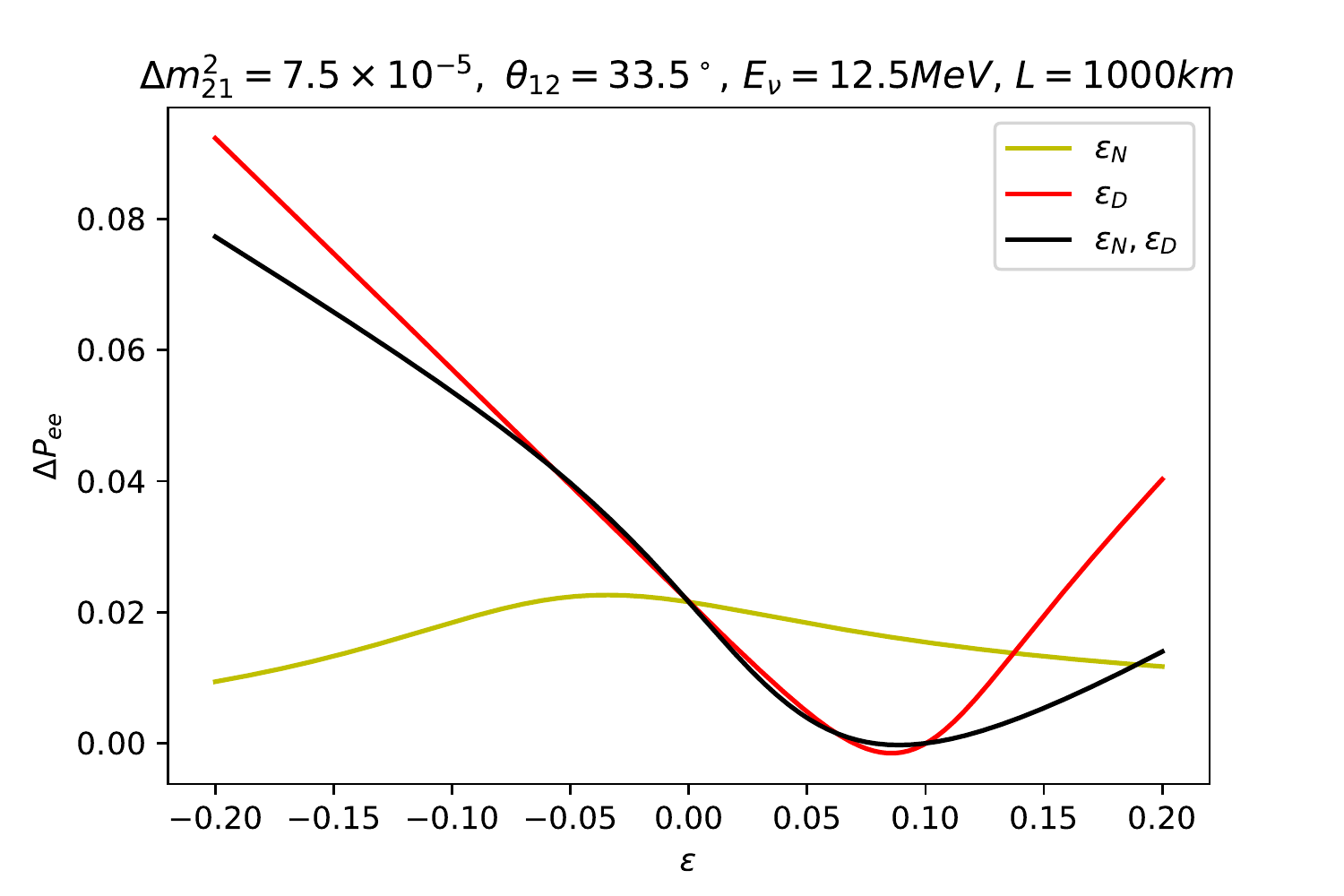}
\hspace{0cm}
\includegraphics[width=0.45\textwidth, height=0.35\textwidth]{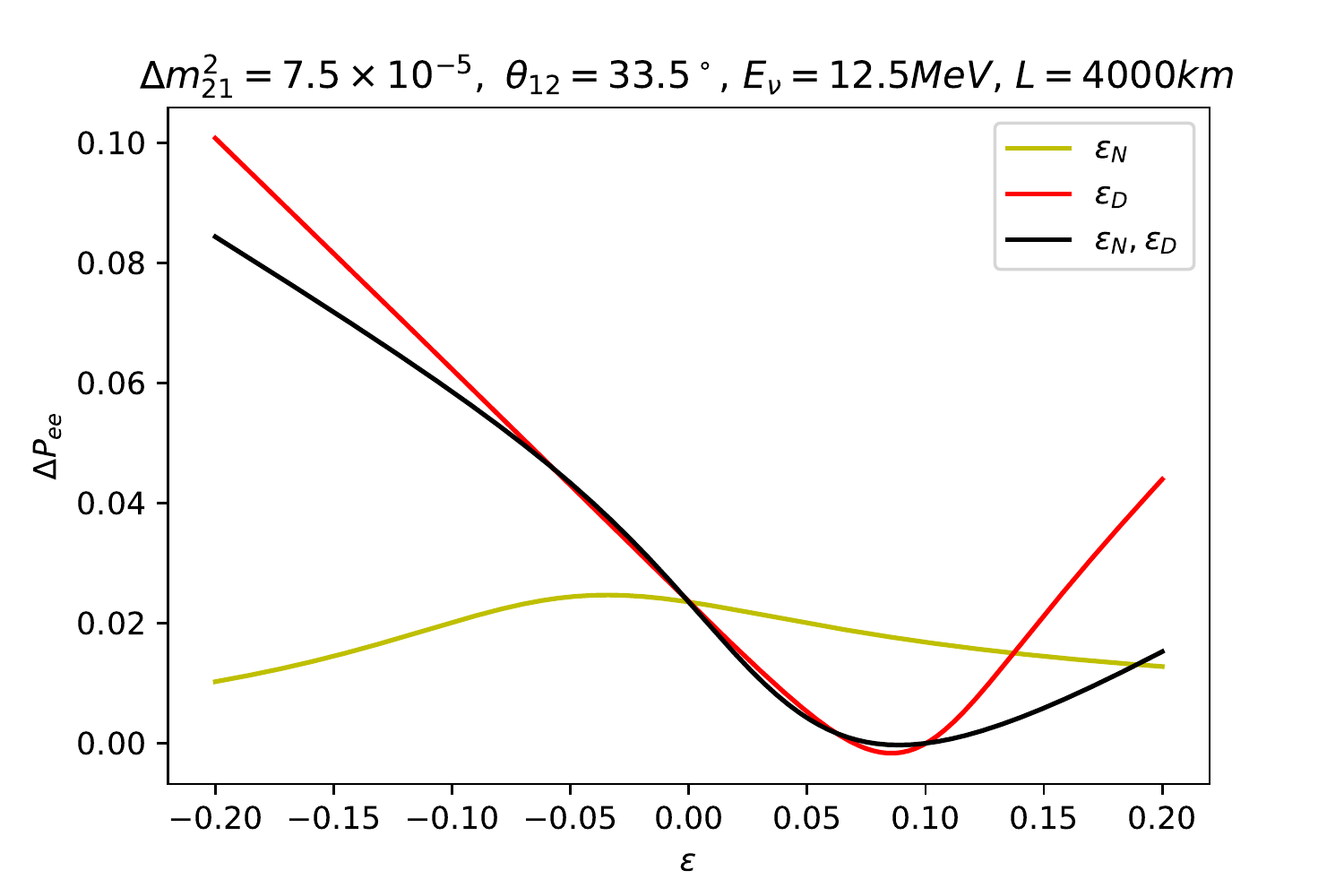}
\hspace{0cm}
\caption[...]{
The differences of survival probability during night and day is plotted versus $\epsilon$ assuming neutrinos with the energy of 12.5 MeV. We have assumed $L=1000$ km and $L=4000$ km in the left and in the right panel, respectively. As it can be seen, $\epsilon_D$ has more significant effect on $\Delta P$.
} \label{Deltapp}
\end{figure}

\begin{figure}[h]
\hspace{0cm}
\includegraphics[width=0.4\textwidth, height=0.35\textwidth]{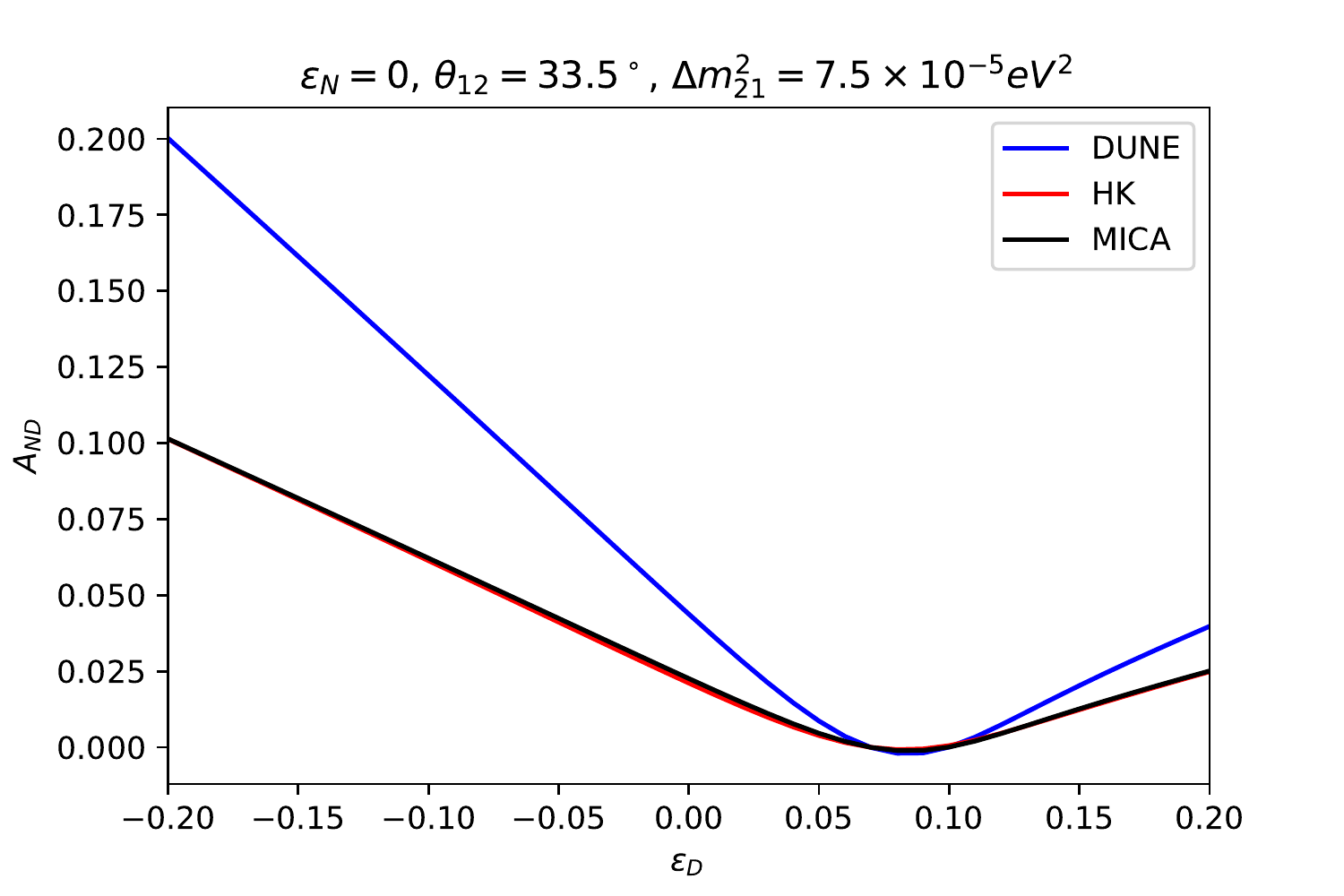}
\hspace{0cm}
\includegraphics[width=0.4\textwidth, height=0.35\textwidth]{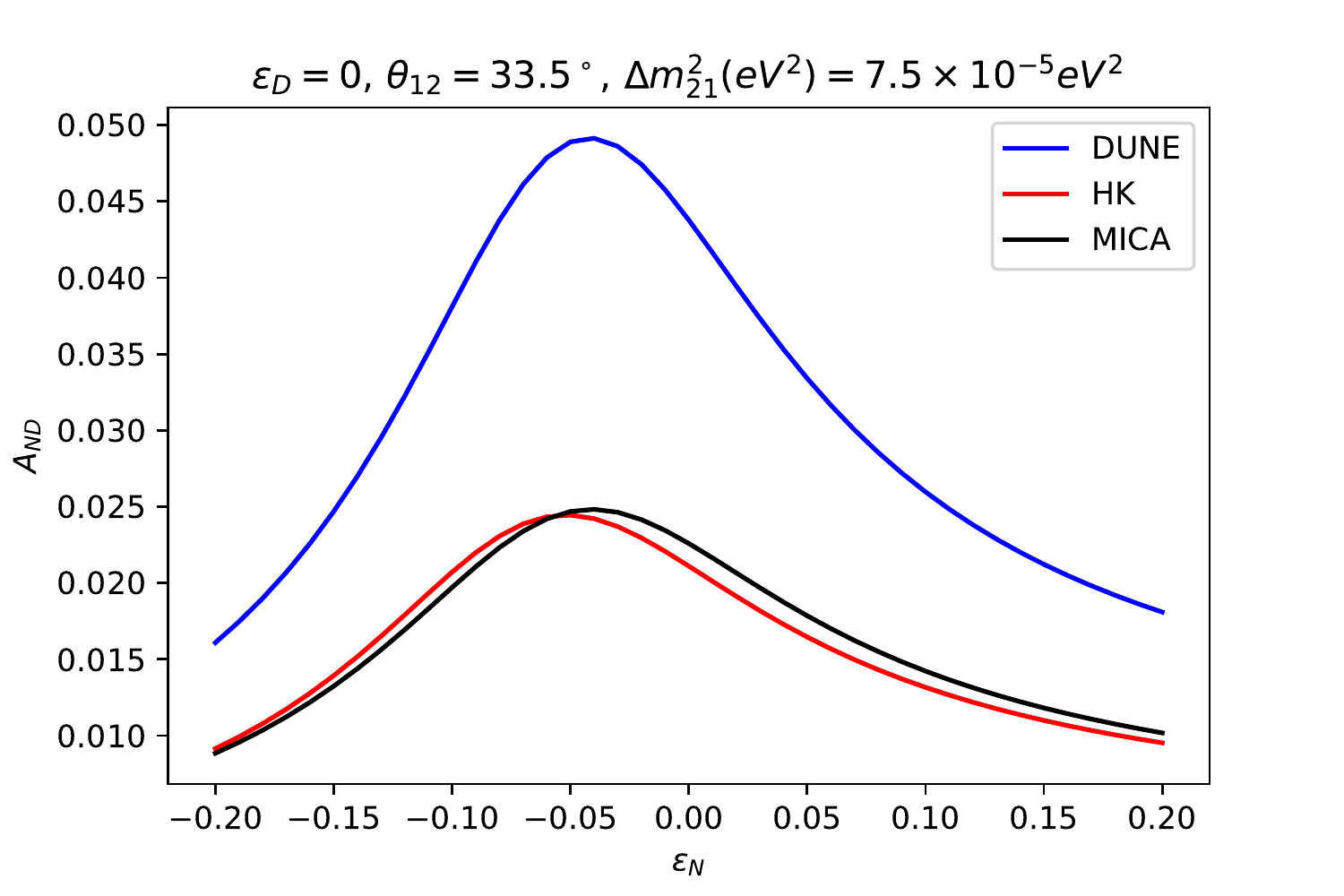}
\hspace{0cm}
\caption[...]{
$A_{ND}$ as a function of $\epsilon$ is plotted for DUNE, HK and MICA experiments.} \label{A_ND}
\end{figure}

\begin{figure}[h]
\hspace{0cm}
\includegraphics[width=0.4\textwidth, height=0.35\textwidth]{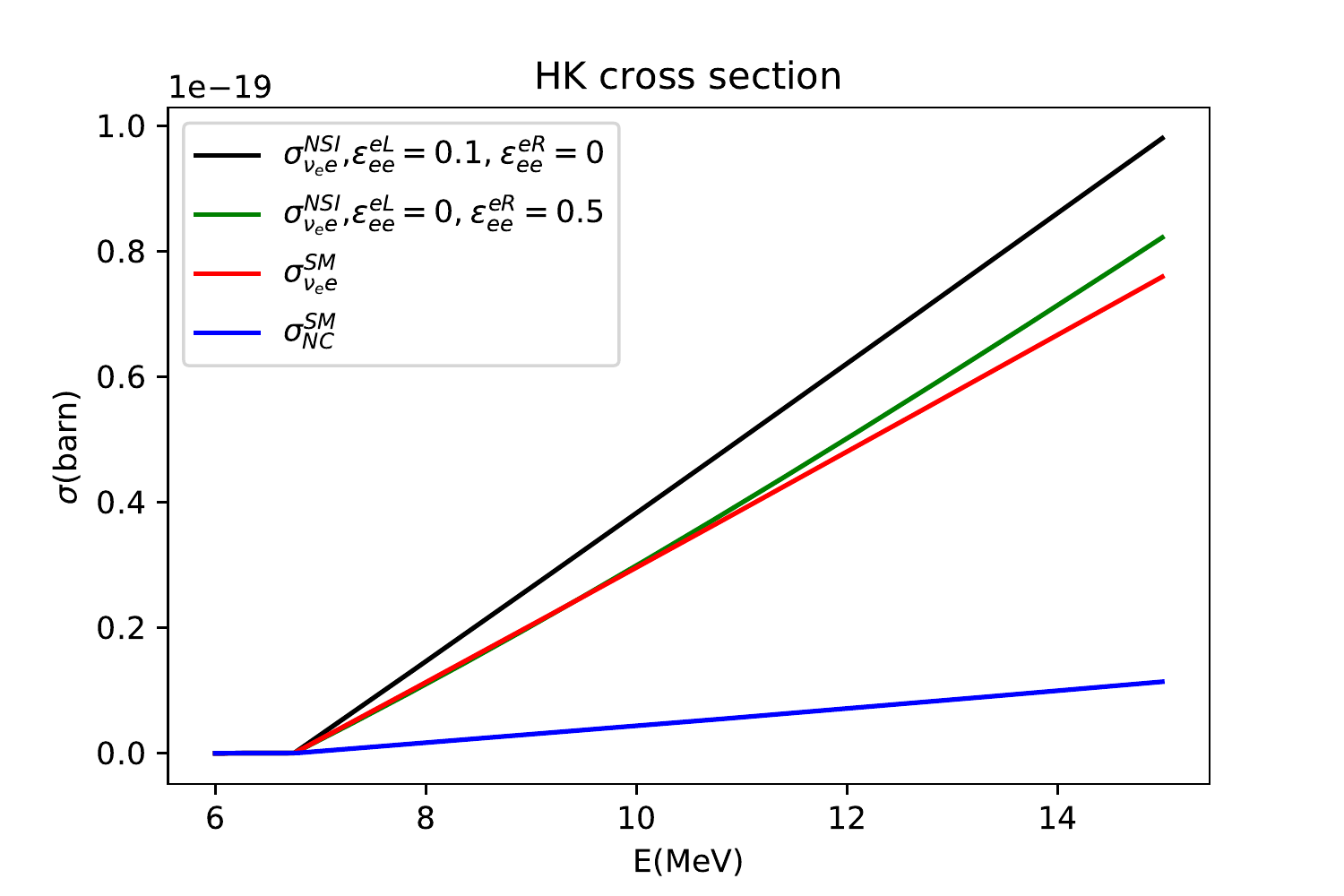}
\hspace{0cm}
\includegraphics[width=0.4\textwidth, height=0.35\textwidth]{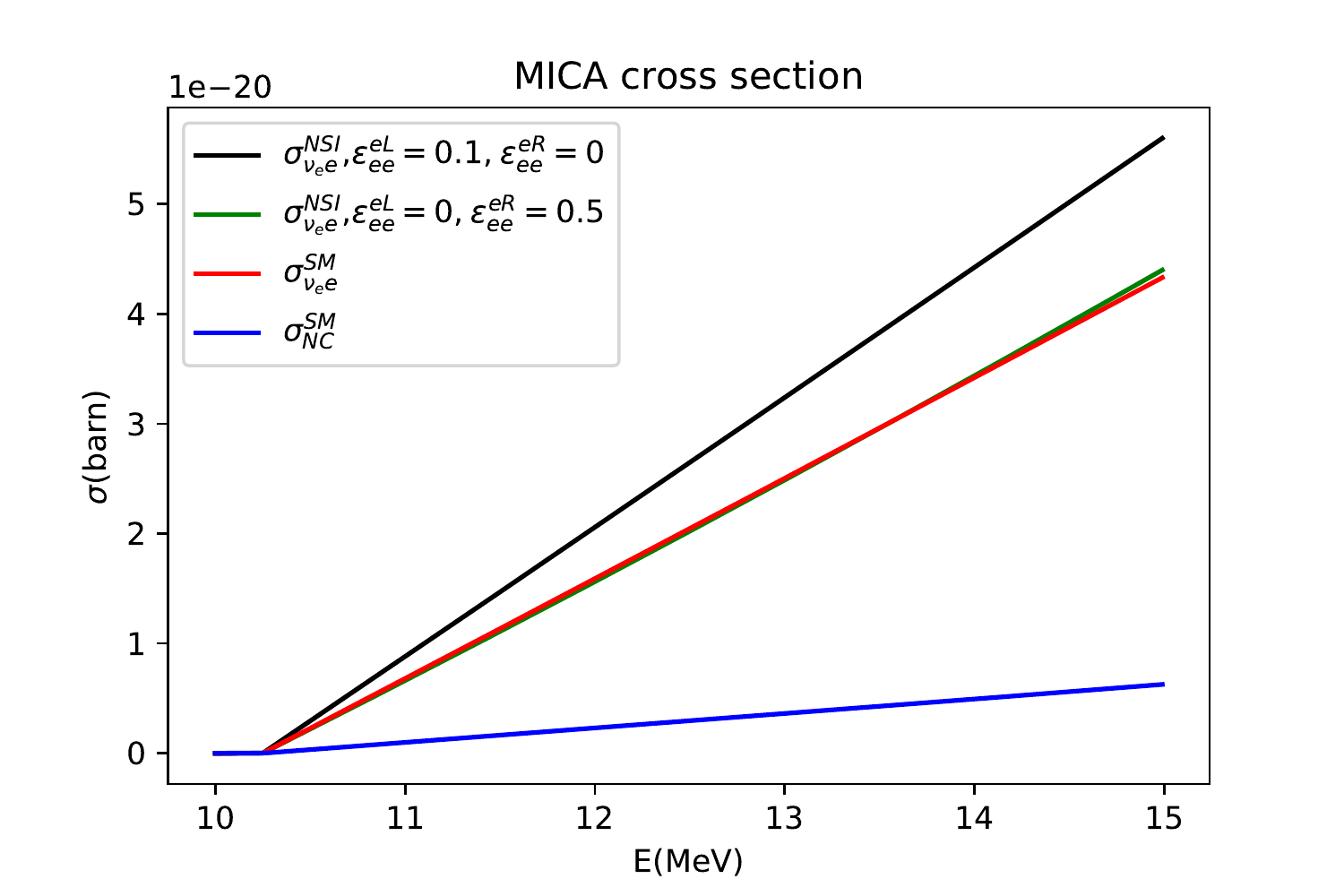}
\hspace{0cm}
\caption[...]{
Total neutrino electron scattering cross section versus the neutrino energy. In the left panel, neutrino electron scattering cross section is plotted for HK experiment with the energy threshold of $6.5$ MeV. The red line is plotted considering standard model cross section for electron neutrinos, the blue line indicates NC cross section for muon neutrinos, the green line is plotted considering NSI for electron neutrino interactions considering large value of $\epsilon_{ee}^{eR}  = 0.5$ and the black line is plotted considering NSI for electron neutrino interactions considering large value of $\epsilon_{ee}^{eL}  = 0.1$. In the right panel, neutrino electron scattering cross section is plotted considering MICA experiment with the energy threshold of $10$ MeV. As it can be seen from the plot, the effect of $\epsilon_{ee}^{eR} $ is as a minor effect.
\label{cross}
}
\end{figure}

In this section, we discuss the main results of our analysis and
discuss the possibility to constrain NSI parameters, considering the future solar neutrino experiments, namely,
DUNE, HK and MICA.

We have plotted $\cos 2 \tilde{\theta}_{12}$ for different values of energy and $\epsilon$ in Fig.~\ref{cos2t_sun1}, considering $\epsilon_{\alpha\beta}^e=\epsilon_{\alpha\beta}^u=\epsilon_{\alpha\beta}^d=\epsilon_{\alpha\beta}$. As it is indicated in the plot, for values of $|\epsilon|$ less than 0.01, the $\epsilon_D$ is indistinguishable from $\epsilon_N$ in solar neutrino oscillation probability. For solar neutrinos with energies more than 10 MeV, there is a degeneracy between $\epsilon_D$ and $\epsilon_N$ for values of $\epsilon$ less than 0.03. In Fig.~\ref{cos2t_sun2}, we have plotted $\cos 2 \tilde{\theta}_{12}$ versus energy considering value of $\epsilon_D$ and/or $\epsilon_N$ is equal 0.1.


\begin{figure}[h]
\hspace{0cm}
\includegraphics[width=0.4\textwidth, height=0.35\textwidth]{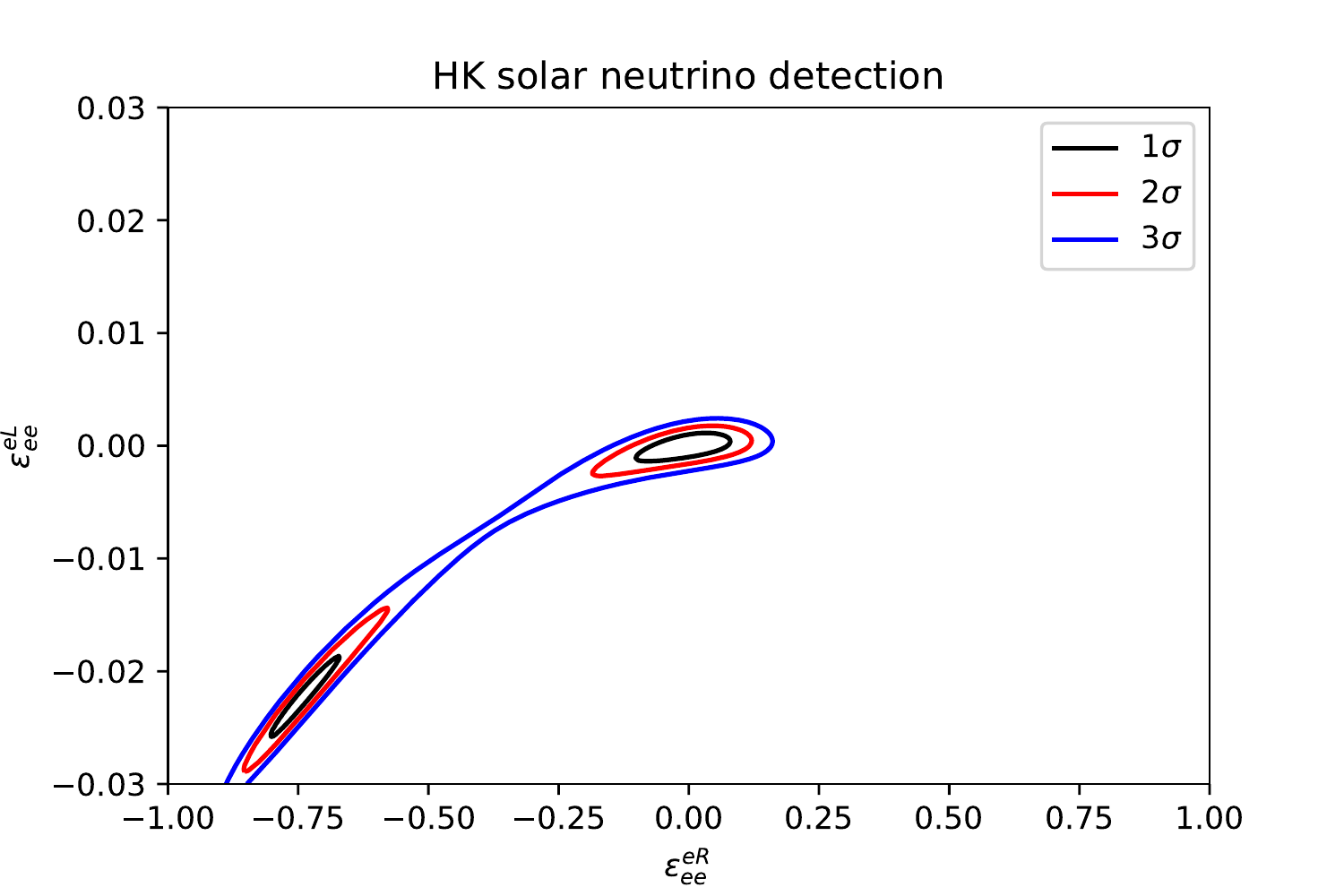}
\hspace{0cm}
\includegraphics[width=0.4\textwidth, height=0.35\textwidth]{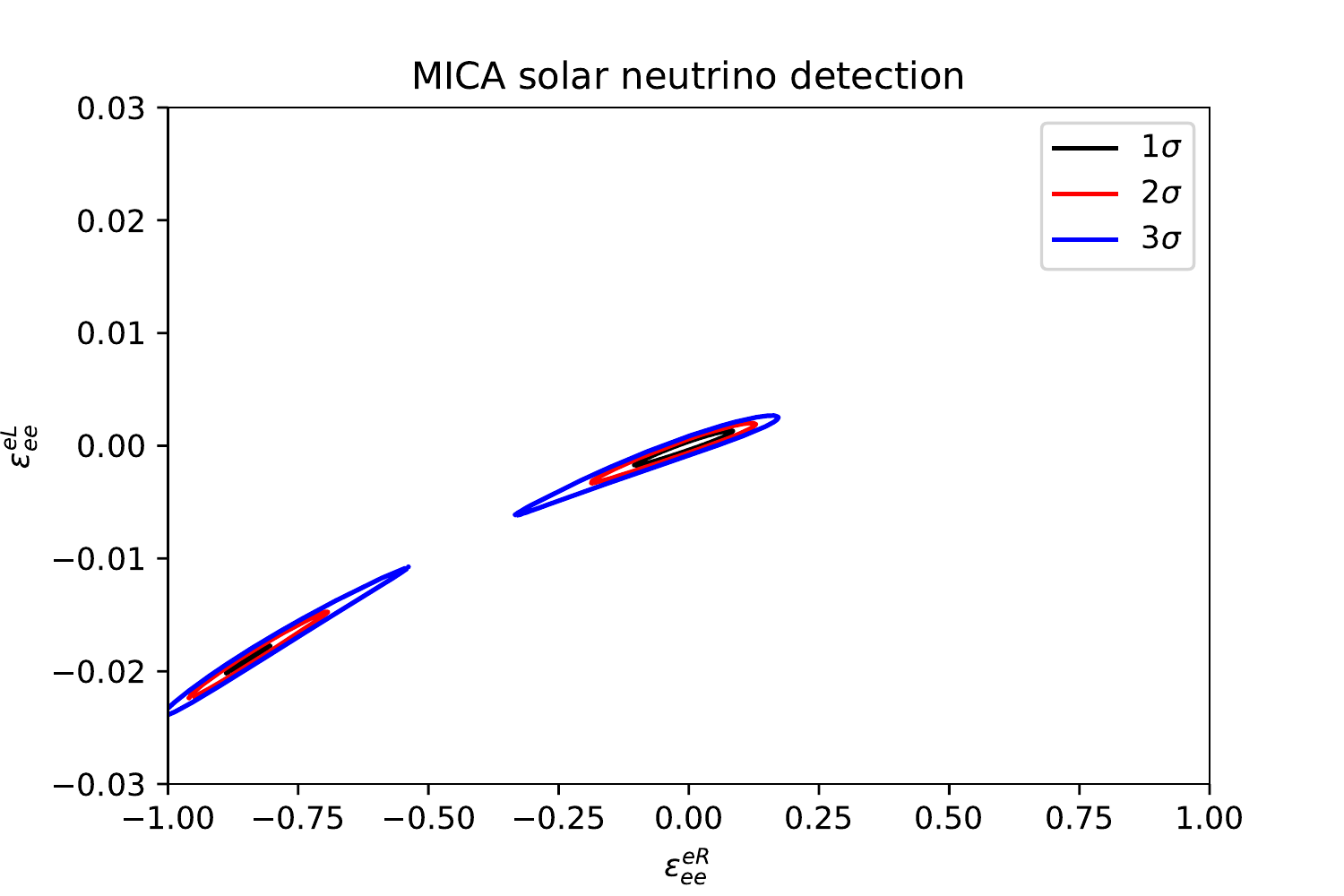}
\hspace{0cm}
\caption[...]{
1$\sigma$, 2$\sigma$ and 3$\sigma$ allowed regions for $\epsilon_{ee}^{eL}$ versus $\epsilon_{ee}^{eR}$ for HK (left) and MICA (right).}\label{eps_LR}
\end{figure}

\begin{figure}[h]
\hspace{0cm}
\includegraphics[width=0.4\textwidth, height=0.35\textwidth]{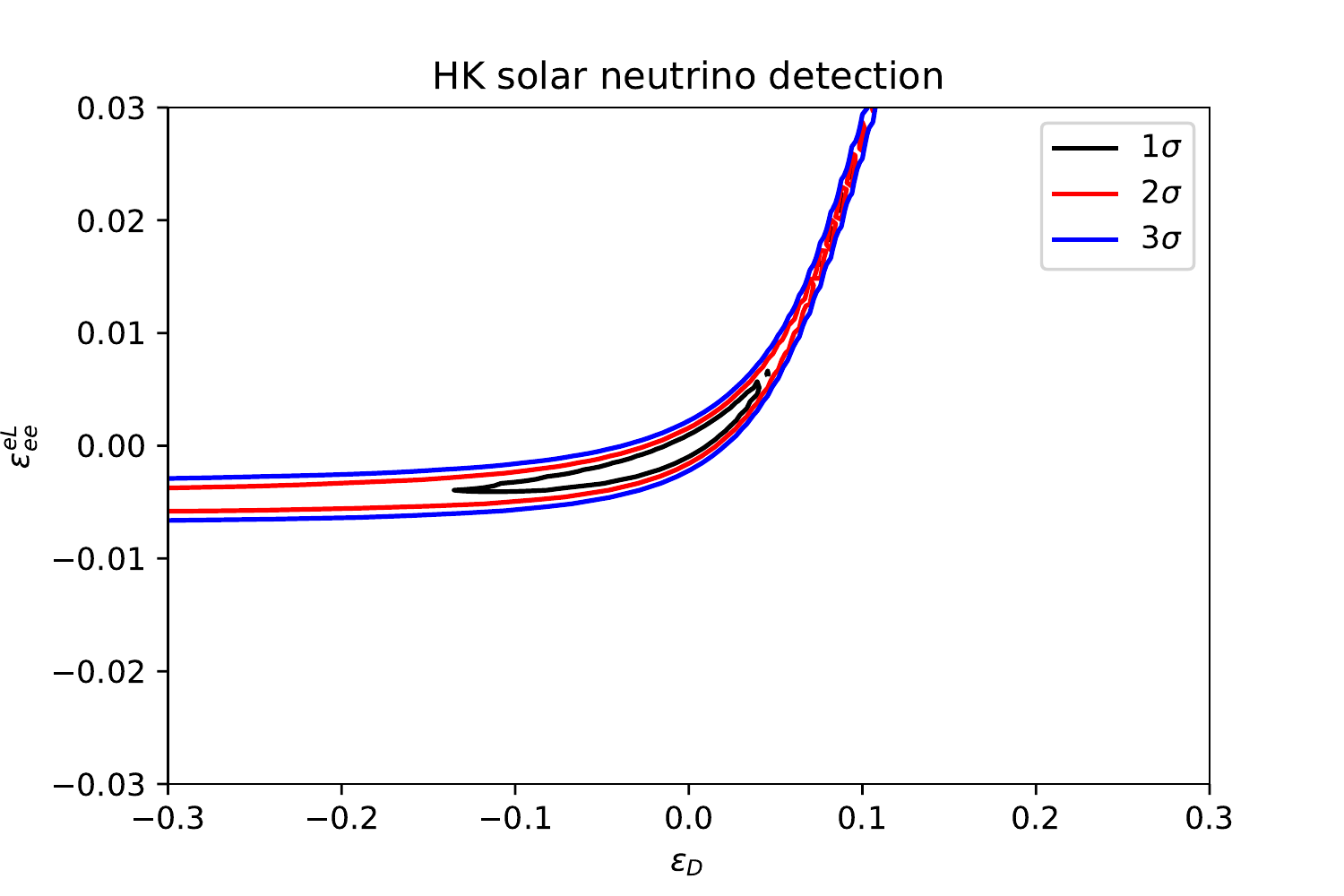}
\hspace{0cm}
\includegraphics[width=0.4\textwidth, height=0.35\textwidth]{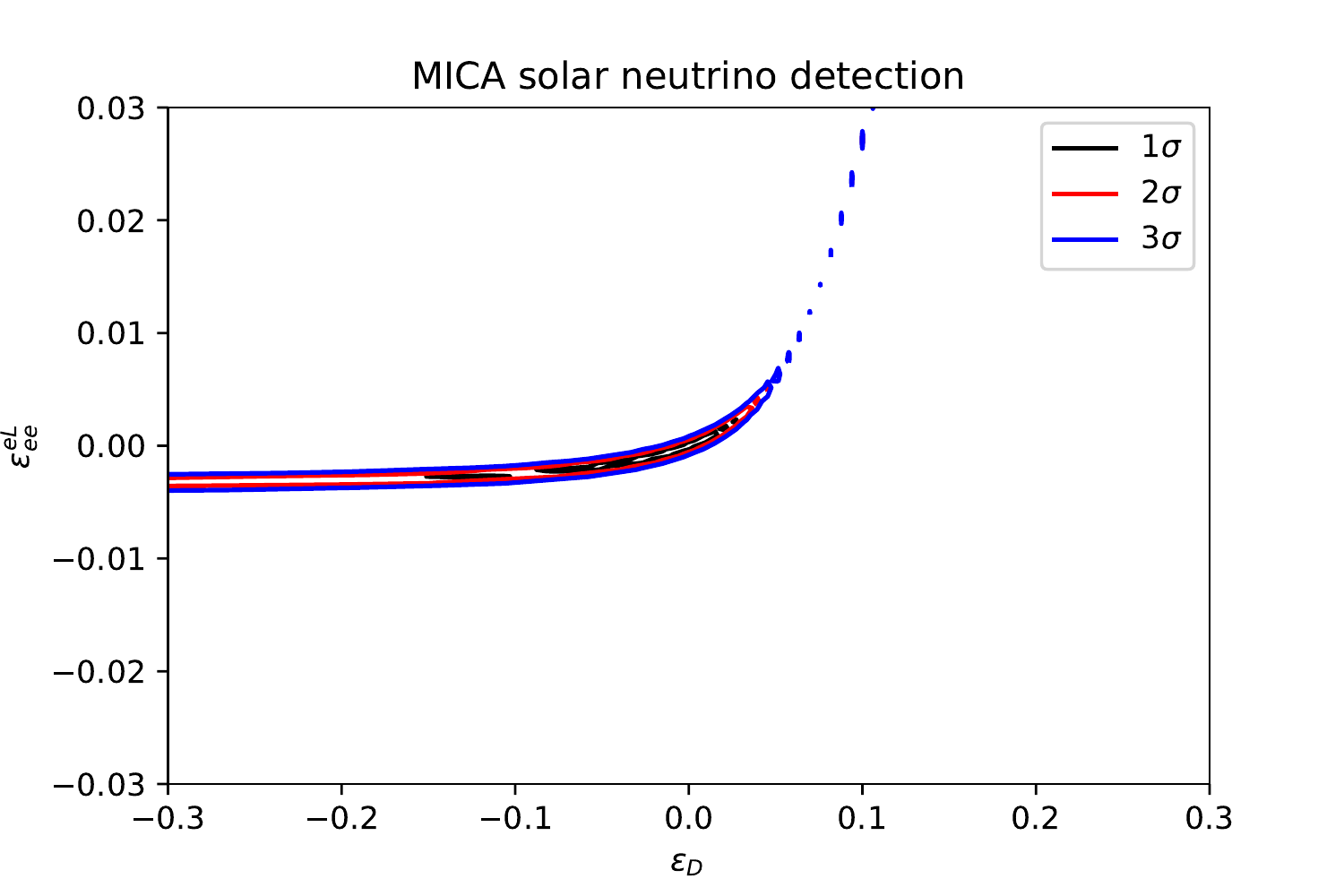}
\hspace{0cm}
\includegraphics[width=0.4\textwidth, height=0.35\textwidth]{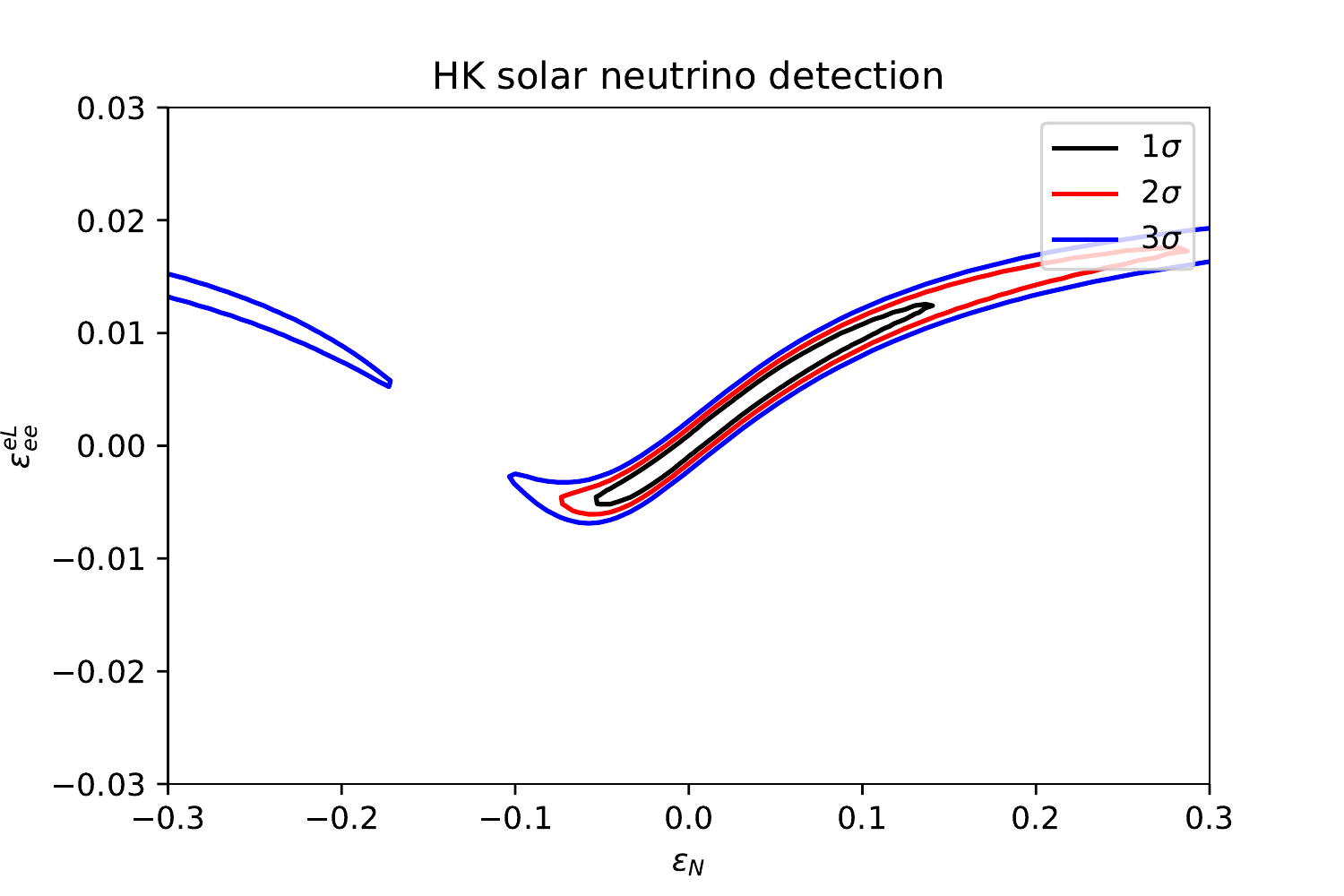}
\hspace{0cm}
\includegraphics[width=0.4\textwidth, height=0.35\textwidth]{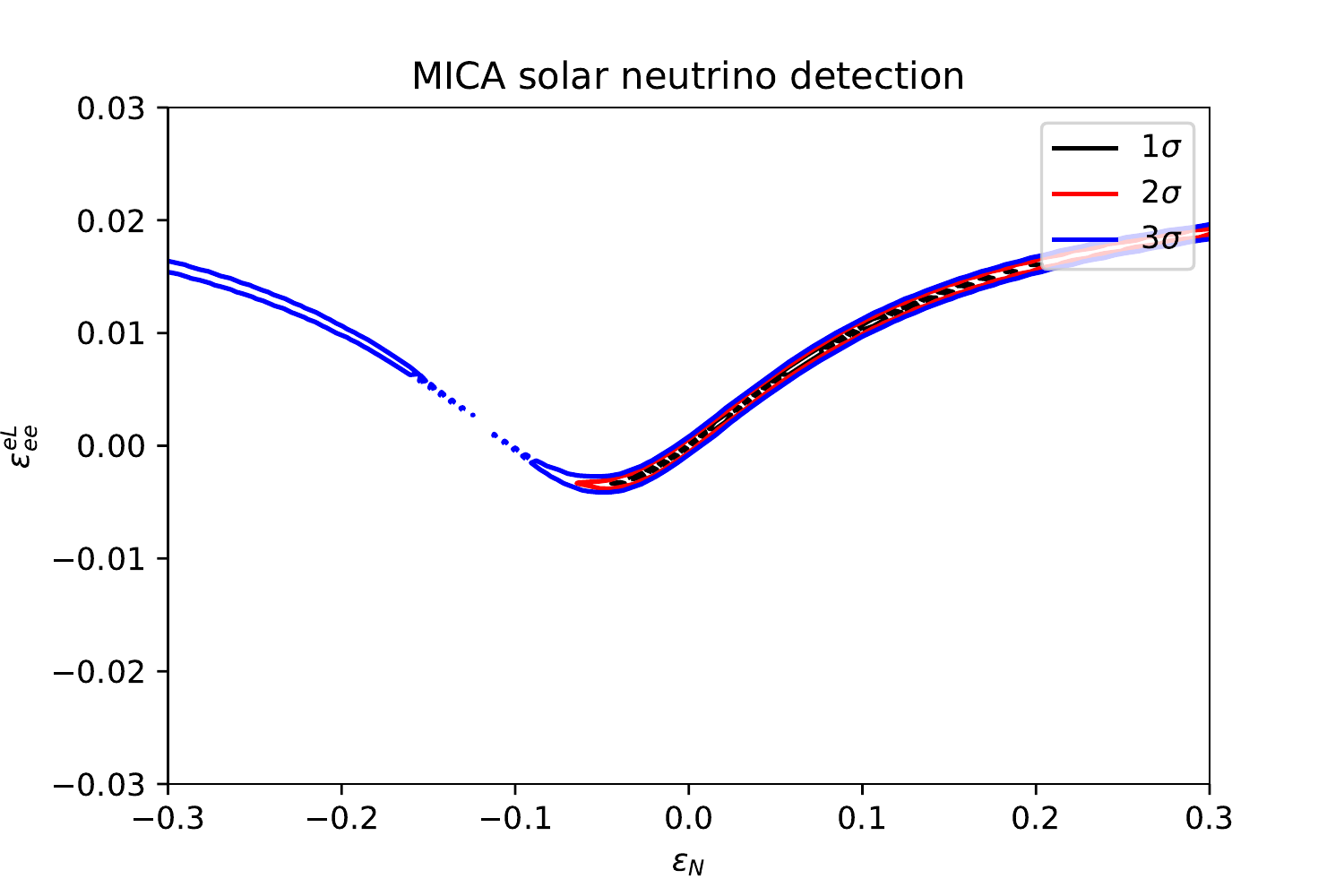}
\caption[...]{
1$\sigma$, 2$\sigma$ and 3$\sigma$ allowed region plotted for simultaneous measurement of $\epsilon_D$ and $\epsilon_{ee}^{eL}$ for HK (upper left panel) and for MICA (upper right panel). The lower panel indicates 1$\sigma$, 2$\sigma$ and 3$\sigma$ allowed region for the simultaneous measurement of $\epsilon_N$ and $\epsilon_{ee}^{eL}$ with HK (lower left) and MICA (lower right).
\label{epsDN_R}
}
\end{figure}

Considering different solar neutrino observatories, DUNE, HK, MICA, and combination of DUNE and HK, the constraints on $\epsilon_D$ and $\epsilon_N$ after ten years of data taking are demonstrated in Fig.~\ref{constraints_eps}. As it is demonstrated the constraints on $\epsilon_D$ and $\epsilon_N$ will be of the order of few 0.01. As discussed before and it is demonstrated in Fig.~\ref{cos2t_sun1}, there will be a degeneracy between $\epsilon_D$ and $\epsilon_N$, and also there is a strong anti-correlation between these two parameters.

We have calculated $\Delta P(E,\eta)$ numerically, considering the PREM model, for different values of energy. Since the peak of the events approximately corresponds to 12.5~MeV, we have demonstrated the results for this value of energy and $L=1000$~km and 4000~km in Fig.~\ref{Deltapp}. As it is obvious the effect of $\epsilon_N$ is distinguishable from the effect of $\epsilon_D$ and the effect of $\epsilon_D$ is more significant on $\Delta P(\eta)$, while in comparison to $\epsilon_D$, $\epsilon_N$ has a minor effect of day-night asymmetry. Considering other values of energy and baseline also leads to similar results.

We have also calculated $\bar{A}_{DN}$ for different values of $\epsilon_D$ and $\epsilon_N$ for three different experiments, DUNE, Hyper Kamiokande, and MICA, for $\theta_{12}=33.5^\circ$ and $\Delta m^2_{21}=7.5\times10^{-5}$ ~$eV^2$. The statistical precision measurement of DUNE, HK, and MICA will be 0.002, 0.001, and 0.0005 at 1$\sigma$ C.L. after 10 years of data taking of solar neutrinos.
The uncertainty of the value of $A_{ND}$ from the uncertainty of $\Delta m^2_{21}$ will be 0.0007, 0.0004, and 0.004 for DUNE, HK, and MICA respectively. Uncertainty of $\theta_{12}$ does not have any significant effect on $A_{ND}$.
As it is indicated in Fig~\ref{A_ND}, the constraints on $|\epsilon_N|$, considering both the statistical precision measurement and systematic uncertainty due to measurement of $\Delta m^2_{21}$, will be 0.014,0.014, and 0.007, and the constraints on $|\epsilon_D|$ are 0.004, 0.004, and 0.002, respectively with DUNE, HK and MICA.

If the future experiments establish a larger day-night asymmetry, similar to the value found by current Super-Kmiokande ($A_{ND}=3.3\pm 1\pm 0.5$ percent) \cite{Abe:2016nxk}, considering that the prediction of best point fit of solar parameters from Kamland is 1.7 percent, such a significant difference cannot be explained by the allowed values of $\epsilon_N$, but it can be explained by the allowed values of $\epsilon_D$. Notice that both $\epsilon_N$ and $\epsilon_D$ will be constrained stringently by the oscillation of neutrinos in the Sun.

Up to this point, in all our previous calculations for examining day-night asymmetry and oscillation of the neutrino in the sun, we have used the standard model cross-section. In Fig. \ref{cross}, total neutrino electron scattering cross-section versus the neutrino energy is demonstrated considering different cases; The red curve is plotted for HK experiment, considering standard model cross-section for electron neutrinos, the blue curve indicates NC cross-section for muon neutrinos, the green line is plotted considering NSI for electron neutrino interactions considering the large value of $\epsilon_{ee}^{eR} = 0.5$ and the black line is plotted considering NSI for electron neutrino interactions considering the large value of $\epsilon_{ee}^{eL} = 0.1$. We have considered the energy threshold of $6.5$ MeV for the HK experiment. In the right panel, the results are indicated for MICA experiment, considering the energy threshold of $10$ MeV. As it can be seen from the plot, the effect of $\epsilon_{ee}^{eR}$ is as a minor effect.

Moreover, it is interesting to investigate the effect of NSI at the detector in the simulated data for these experiments.
We have demonstrated the potential of HK and MICA to constrain $\epsilon_{ee}^{eL}$ versus $\epsilon_{ee}^{eR}$ in Fig~\ref{eps_LR}. As it can be seen from the plot, the constraint on $\epsilon_{ee}^{eL}$ is very stringent while in the case of $\epsilon_{ee}^{eR}$ is not.

The uncertainty of $\epsilon_{ee}^{eL}$ and $\epsilon_{ee}^{eR}$ can affect HK and MICA measurement of $\epsilon_D$ and $\epsilon_N$ from the oscillation of the neutrino in sun. In the case of DUNE, since the neutrino detection is via charged current interaction, adding the NC interaction does not affect the cross-section. However, if we consider charged current NSI, the NSI parameters will be constrained stringently by DUNE near detector down to the order 0.001 as studied in ref. \cite{Bakhti:2016gic}; Thus, considering CC NSI does not affect the cross-section of DUNE and in consequence measurement of $\epsilon_D$ and $\epsilon_N$.

Notice that uncertainties of the cross-section and the flux do not affect day-night asymmetry. It can be seen from Eq. \ref{day}
where the uncertainties enter in both the denominator and in the numerator; Thus, the uncertainties of $\epsilon_{ee}^{eL}$ and $\epsilon_{ee}^{eR}$ does not affect the measurement of $\epsilon_N$ and $\epsilon_D$ in the day-night asymmetry.

Since $\epsilon_{ee}^{eL}$ has a more significant impact of neutrino electron elastic scattering cross-section than $\epsilon_{ee}^{eR}$,
to find the impact of cross-section uncertainty on $\epsilon_D$ and $\epsilon_N$ measurements from the oscillation of the neutrinos in the sun, we study the potential of HK and MICA simultaneous measurement of $\epsilon_{ee}^{eL}$ and $\epsilon_D$ and simultaneous measurement of $\epsilon_{ee}^{eL}$ and $\epsilon_N$. The results are demonstrated in Fig.~\ref{epsDN_R}. As it is shown, $\epsilon_{ee}^{eL}$ uncertainty has a huge impact on $\epsilon_D$ and $\epsilon_N$ measurement, with HK and MICA from neutrino oscillation in the sun. However, as it is explained before, the constraints from day-night asymmetry are not affected.

As it is indicated in Fig. 3, DUNE will constrain $\epsilon_D$ and $\epsilon_N$ down to 0.01 with 1$\sigma$, and $\epsilon_N$ and $\epsilon_D$ will be constrained by Day-Night asymmetry down to 0.01 and 0.004, respectively; Thus, $\epsilon_{ee}^{eL}$ will be constrained down to the order of $0.001$, combining of DUNE and HK results.

\section{Summary \label{sum}}

We studied the sensitivities to NSI in the proposed next generation solar neutrino
observatories DUNE, HK and MICA.
The  reactor experiment JUNO will be able to measure $\Delta m^2 _{12}$ and $\theta_{12}$ with less than one percent precision.
On the other hand, having relatively low energy, JUNO is not sensitive to the standard and non-standard matter effects.
To study the effect of NSI parameters on the experimental performance, we
considered this possible precise measurement of $\Delta m^2 _{12}$ and $\theta_{12}$ by JUNO. We also assumed same NSI couplings for electron, up and down quarks ($\epsilon_{\alpha\beta}^e=\epsilon_{\alpha\beta}^u=\epsilon_{\alpha\beta}^d$).

Considering neutrino oscillation in the Sun, we demonstrated the constrains on $\epsilon_D$ and $\epsilon_N$ down to order of 0.01 at 3 $\sigma$ C.L. after ten years of data taking in Fig.~\ref{constraints_eps}, for future experiments DUNE, HK and the proposed MICA experiment in addition to the combination of DUNE and HK. We found that for the values of $|\epsilon|$ less than the order of 0.01, the $\epsilon_D$ is indistinguishable from $\epsilon_N$ in solar neutrino oscillation probability as indicated in Fig.~\ref{cos2t_sun1};
We further studied the day-night asymmetry parameters for three different experiments, DUNE, HK and MICA. As we indicated in Fig.~\ref{A_ND}, $\epsilon_D$ is significantly affected on day-night asymmetry while $\epsilon_N$ is not.
We discussed that for the case of the larger value of day-night asymmetry that may be established by future experiments, the allowed values of $\epsilon_N$ cannot explain such a huge difference while it can be explained by the allowed values of $\epsilon_D$.

Besides, we studied the effect of NSI at the detector for the simulated
data for these experiments. We demonstrated the potential of HK and MICA to constrain NSI.
Our results show that while $\epsilon_{ee}^{eR}$ is weakly constrained,
the constraint on $\epsilon_{ee}^{eL}$ is very stringent. We further studied the potential of HK and MICA simultaneous measurement of $\epsilon_{ee}^{eL}$ and $\epsilon_D$ and simultaneous measurement of $\epsilon_{ee}^{eL}$ and $\epsilon_N$. We found that
for HK and MICA experiments, $\epsilon_{ee}^{eL}$ uncertainty has a significant effect on $\epsilon_D$ and $\epsilon_N$ measurement considering neutrino oscillation in the Sun.

 \subsection*{Acknowledgments}
 The authors are grateful to   E. Fernandez Martinez  for useful discussion.
This project has received funding from the European Union's Horizon 2020 research and innovation programme under the Marie Sk\l{}odowska-Curie grant agreement No.~674896 and No.~690575. We are very grateful to the IFT institute, UAM
University   for warm and generous hospitality during this project. P.B thanks Iran Science Elites Federation Grant No. 11131.


\end{document}